\documentclass[twocolumn,epjc3]{svjour3}
\usepackage{graphics}
\usepackage{microtype}
\usepackage{bbm}
\usepackage{amsmath}
\usepackage{mathtools}
\usepackage{caption}
\usepackage{subcaption}
\usepackage{amssymb}
\usepackage{mathrsfs}       
\usepackage[pdftex]{graphicx}
\usepackage{pgfplots}
\pgfplotsset{width=10cm,compat=1.9}
\usepackage{xargs}       
\usepackage{wasysym}
\usepackage{engrec}
\usepackage{enumerate}
\usepackage{appendix}
\usepackage{empheq}
\usepackage{float}
\usepackage{stmaryrd}
\usepackage[parfill]{parskip}
\usepackage[pdftex,breaklinks]{hyperref}
\usepackage{titletoc}
\usepackage{makecell}
\usepackage{lscape}
\usepackage{algorithm}
\usepackage{algorithmicx}
\usepackage{tensor}
\usepackage{algpseudocode}
\usepackage{blkarray}
\usepackage{multirow}
\usepackage{bigstrut}
\usetikzlibrary{plotmarks}

\newcommand{\ket}[1]{|#1\rangle}
\newcommand{\bra}[1]{\langle#1|}

\newcommand{\nucl}[2]{\({}^{#2}\mathrm{#1}\)}

\newenvironment{customlegend}[1][]{%
    \begingroup
    \csname pgfplots@init@cleared@structures\endcsname
    \pgfplotsset{#1}%
}{%
    \csname pgfplots@createlegend\endcsname
    \endgroup
}%
\def\addlegendimage{\csname pgfplots@addlegendimage\endcsname}

\definecolor{mygreen}{RGB}{21, 158, 10}
\begin{document}
\title{Rooting the EDF method into the ab initio framework}
\subtitle{PGCM-PT formalism based on MR-IMSRG pre-processed Hamiltonians}
\author{T. Duguet\thanksref{ad:saclay,ad:kul} \and J.-P. Ebran\thanksref{ad:dam,ad:fakedam}  \and M. Frosini\thanksref{ad:saclay,ad:des} \and H.~Hergert\thanksref{ad:msu1,ad:msu2} \and V. Som\`a\thanksref{ad:saclay} }

\institute{
\label{ad:saclay}
IRFU, CEA, Universit\'e Paris-Saclay, 91191 Gif-sur-Yvette, France 
\and
\label{ad:kul}
KU Leuven, Department of Physics and Astronomy, Instituut voor Kern- en Stralingsfysica, 3001 Leuven, Belgium 
\and
\label{ad:dam}
CEA, DAM, DIF, 91297 Arpajon, France
\and
\label{ad:fakedam}
Universit\'e Paris-Saclay, CEA, Laboratoire Mati\`ere en Conditions Extr\^emes, 91680, Bruy\`eres-le-Ch\^atel, France
\and
\label{ad:des}
CEA, DES, IRESNE, DER, SPRC, LEPh,
13115 Saint-Paul-lez-Durance, France
\and
\label{ad:msu1}
Facility for Rare Isotope Beams, Michigan State University, East Lansing, MI 48824-1321, USA
\and
\label{ad:msu2}
Department of Physics and Astronomy, Michigan State University, East Lansing, MI 48824-1321, USA
}

\date{Received: \today{} / Revised version: date}

\maketitle
%
%
\begin{abstract}
Recently, ab initio techniques have been successfully connected to the traditional valence-space shell model. In doing so, they can either explicitly provide ab initio shell-model effective Hamiltonians or constrain the construction of empirical ones. In the present work, the possibility to follow a similar path for the nuclear energy density functional (EDF) method is analyzed. For this connection to be actualized, two theoretical techniques are instrumental: the recently proposed ab initio PGCM-PT many-body formalism and the MR-IMSRG pre-processing of the nuclear Hamiltonian. Based on both formal arguments and numerical results, possible new lines of research are briefly discussed, namely to compute ab initio EDF effective Hamiltonians at low computational cost, to constrain empirical ones or to produce them directly via an effective field theory that remains to be invented.
\end{abstract}

\section{Introduction}
\label{intro}

As of today, the main benefit of ab initio calculations is to rely on a {\it systematic} approach to both the input Hamiltonian and the solutions of Schr{\"o}dinger's equation in the full A-body Hilbert space ${\cal H}_{\text{A}}$~\cite{Hergert20}. This feature provides a way to improve the calculation step by step and to deliver associated uncertainties along the way. In recent years, ab initio calculations have managed to enter the domain of mid-mass nuclei thanks to approximation schemes whose numerical cost scales polynomially with the system size~\cite{Hagen14,Hergert16,Soma20b,Tichai20}. Still, the high polynomial cost currently limits the target accuracy to a few $\%$ and/or the target nucleus to mass $\text{A} \lesssim 130$. 

In contrast, the main benefit of the energy density functional (EDF) method relates to a low polynomial cost associated with an a priori restriction of the explicit nucleon dynamics to a small subspace of ${\cal H}_{\text{A}}$~\cite{Bender03,10.1088/2053-2563/aae0ed}. This scheme makes large-scale calculations over the entire nuclear chart possible on a short time scale. However, the reduction to a subspace of ${\cal H}_{\text{A}}$ has been accompanied so far by an empirical formulation of the effective Hamiltonian at play. The current ad hoc character of the EDF method thus makes impossible to systematically test and improve associated predictions.

In this context, a legitimate objective is to build an EDF in a top-down manner, by rooting the EDF method into the ab initio theoretical scheme (for previous works of various flavours towards this objective see, e.g., Refs.~\cite{Duguet15,Grasso16,Salvioni20,Marino21}). The long-term objective is to make the EDF method {\it systematic} in the sense described above while maintaining its key characteristic, i.e. the low numerical cost authorizing routine calculations over the nuclear mass table. In order to achieve this goal, one must have at hand 
\begin{enumerate}
\item an ab initio many-body method whose formulation allows an explicit connection to the subspace reduction at play in the EDF method,
\item an explicit formulation of the ab initio effective, nucleus-dependent, Hamiltonian at play in that subspace,
\item a way to systematically build that effective, nucleus-dependent, Hamiltonian without actually paying the full price of its explicit ab initio construction for each system under study.
\end{enumerate}
The goal of the present paper is to demonstrate, both formally and numerically, that requirements 1. and 2. above are indeed fulfilled today. This important observation opens up a research program whose goal will be to realize point 3. 

The paper is organized as follows. While Sec.~\ref{sec:1} briefly introduces the main characteristics of ab initio and EDF theoretical schemes, Sec.~\ref{sec:2} proceeds to a numerical comparison in $^{20}$Ne, i.e. in a non-trivial doubly open-shell system. Because of the apparent irreconcilable character of ab initio and EDF results, Sec.~\ref{sec:3} adds the key layer to the discussion constituted by the pre-processing of the Hamiltonian that, as demonstrated numerically, constitutes the missing link to reconcile the two theoretical schemes. Based on this analysis, several ideas to design a {\it systematic} EDF method in the future are briefly elaborated on in Sec.~\ref{sec:4}.

\section{Formal connection}
\label{sec:1}

The goal of quantum many-body methods is to deliver many-body observables through a relevant {\it theoretical scheme} based on (i) an appropriate choice of degrees of freedom, (ii) a formulation of their elementary interactions and (iii) the resolution of the appropriate dynamical equation delivering the observables of interest out of these degrees of freedom and interactions.

\subsection{Ab initio PGCM-PT many-body formalism}
\label{abinitio}

\subsubsection{Theoretical scheme}

The so-called {\it ab initio}\footnote{This naming is inappropriate given that the associated theoretical scheme does not start ``from the beginning", i.e. from the most elementary set of degrees of freedom, if any, but simply denotes a meaningful connection to the underlying effective field theory of the strong force, quantum chromodynamics (QCD). It rather constitutes an ``in medias res" scheme as any other effective (field) theory~\cite{Bontems2021}.} description of nuclear systems is nothing but the implementation of the pion-full (or pion-less) effective field theory (EFT) in the A-nucleon sector~\cite{Epelbaum:2008ga,Machleidt:2020vzm,vanKolck:2020plz,vanKolck:2020llt}. Taking chiral EFT ($\chi$EFT) as an example, the theoretical scheme corresponds to describing a nuclear system as a collection of (i) A non-relativistic point-like nucleons (ii) interacting via the exchange of pions and short-range contact operators in all possible ways allowed by the symmetries of QCD. In Weinberg's power counting, observables are obtained (iii) by solving the non-relativistic A-body  Schr{\"o}dinger's equation based on the Hamiltonian $H$ generated in step (ii) at a given order in the $\chi$EFT expansion, i.e.
\begin{align}
H | \Psi^{\sigma}_{\mu} \rangle &=  E^{\tilde{\sigma}}_{\mu} \, | \Psi^{\sigma}_{\mu} \rangle \, . \label{schroedevolstate}
\end{align}
Here $\mu$ denotes a principal quantum number whereas \(\sigma\equiv(\text{J} \text{M} \Pi \text{N} \text{Z})\equiv (\tilde{\sigma}\text{M})\) collects the set of symmetry quantum numbers labelling the many-body states, i.e. the angular momentum J and its projection M, the parity $\Pi=\pm 1$ as well as neutron N and proton Z numbers (A=N+Z). The $M$-independence of the eigenenergies $E^{\tilde{\sigma}}_{\mu}$ and the symmetry quantum numbers carried by the eigenstates are a testimony of the symmetry group 
\begin{align}
\text{G}_{H} \equiv \{R(\theta), \theta \in  D_{\text{G}}\}
\end{align}
of the Hamiltonian, i.e.,
\begin{align}
[H,R(\theta)]=0 \, , \, \forall \theta \, ,
\end{align}
which plays a key role in the present context~\cite{Frosini:2021fjf}.

\subsubsection{Nuclear Hamiltonian}

Built at an order in the $\chi$EFT expansion where, e.g., four- and higher-body interactions vanish, the second-quantized nuclear Hamiltonian reads in an arbitrary basis of the one-body Hilbert space ${\cal H}_1$ as
\begin{align}
    H \equiv & \,\, T + V + W  \notag\\
    \equiv & \,\, \frac{1}{(1!)^2} \sum_{\substack{a_1\\b_1}} t^{a_1}_{b_1} \, A^{a_1}_{b_1} \notag \\
    &\,\,+\frac{1}{(2!)^2} \sum_{\substack{a_1a_2\\b_1b_2}} v^{a_1a_2}_{b_1b_2} \, A^{a_1a_2}_{b_1b_2} \notag \\
    &\,\,+\frac{1}{(3!)^2}  \sum_{\substack{a_1a_2a_3\\b_1b_2b_3}} w^{a_1a_2a_3}_{b_1b_2b_3} \, A^{a_1a_2a_3}_{b_1b_2b_3} \, .  \label{originalH}
\end{align}
In Eq.~\ref{originalH}
\begin{equation} 
A^{a_1\cdots a_n}_{b_1\cdots b_n}\equiv c^\dag_{a_1}\cdots c^\dag_{a_n}c_{b_n}\cdots c_{b_1}
\end{equation}
denotes a string of $n$ one-particle creation and $n$ one-particle annihilation operators. While $t^{a_1}_{b_1} $ defines matrix elements of the kinetic energy, $v^{a_1a_2}_{b_1b_2}$ and $w^{a_1a_2a_3}_{b_1b_2b_3}$ characterize anti-symmetrized matrix elements of two- and three-body interactions, respectively. The nuclear Hamiltonian is thus effectively defined through a set of mode-$2k$ tensors associated with each $k$-body operator involved, the set being limited to $k\leq 3$ in the above example.

\subsubsection{VSRG pre-processing}

An important aspect of the present discussion regards the possibility to pre-process the Hamiltonian via unitary similarity renormalization group (SRG) transformations. For now, the {\it nucleus-independent} pre-processing based on a vacuum SRG (VSRG) transformation $U(\lambda)$ is considered in order to decouple a low-energy subspace from higher momentum modes~\cite{Bogner:2009bt,PhysRevLett.107.072501,PhysRevC.90.024325}. Starting from $H$ in Eq.~\eqref{originalH}, the transformed Hamiltonian reads as
\begin{align}
    H(\lambda) \equiv &  \,\,  U(\lambda)HU^{\dagger}(\lambda)  \notag\\
    \equiv&  \,\, T + V(\lambda) + W(\lambda) + \ldots  \notag\\
    \equiv&  \,\,  \frac{1}{(1!)^2} \sum_{\substack{a_1\\b_1}} t^{a_1}_{b_1} \, A^{a_1}_{b_1} \notag \\
    & \,\, +\frac{1}{(2!)^2} \sum_{\substack{a_1a_2\\b_1b_2}} v^{a_1a_2}_{b_1b_2}(\lambda) \, A^{a_1a_2}_{b_1b_2} \notag \\
    & \,\, +\frac{1}{(3!)^2}  \sum_{\substack{a_1a_2a_3\\b_1b_2b_3}} w^{a_1a_2a_3}_{b_1b_2b_3}(\lambda) \, A^{a_1a_2a_3}_{b_1b_2b_3}\notag \\
    & \,\, +\ldots \, ,  \label{vsrgH}
\end{align}
such that $k$-body operators with $k>3$ are induced by the transformation. Correspondingly, the tensors defining $H(\lambda)$ acquire an explicit dependence on the parameter $\lambda$  parametrizing the transformation. The benefit of the pre-processing is to accelerate the convergence of the subsequent solving of Eq.~\eqref{schroedevolstate}~\cite{Bogner:2009bt,PhysRevLett.107.072501,PhysRevC.90.024325}. This advantage must not be overtaken by more significant drawbacks, i.e. the VSRG transformation is to be stopped before induced four- and higher-body operators become too large. As a result, these induced operators are eventually discarded such that the end Hamiltonian retains the same formal expression and rank\footnote{The Hamiltonians $H$ and $H(\lambda)$ would be unitarily equivalent, i.e. they would deliver identical many-body observables through Eq.~\eqref{schroedevolstate}, if the unitarity of the VSRG transformation was not violated by the omission of four- and higher-body forces into $H(\lambda)$, which thus constitutes an approximation.} as the original $H$ in Eq.~\eqref{originalH}. For simplicity, the $\lambda$-dependence of the end result $H(\lambda)$ is omitted below. 

\subsubsection{Rank-reduction of the Hamiltonian}
\label{rankred}

The computational cost of ab initio nuclear structure calculations is rendered particularly acute by three-nucleon interactions. Consequently, ab initio calculations of mid-mass nuclei are (currently) performed while approximating three-nucleon operators in $H$ in terms of effective, i.e. system-dependent, zero-, one- and two-nucleon operators. This is achieved in doubly closed-shell nuclei via the normal-ordered two-body approximation procedure~\cite{RoBi12} whereas the rank-reduction generalization technique proposed recently in Ref.~\cite{Frosini:2021tuj} allows one to do so in both closed and open-shell nuclei at low computational cost. Starting from a Hamiltonian pre-processed via VSRG transformations, these approximations induce errors below $2-3\%$ across a large range of nuclei, observables and many-body methods.

In this context, the VSRG-transformed Hamiltonian $H$ is approximated by the {\it nucleus-dependent} Hamiltonian\footnote{The Hamiltonians $H$ and $\bar{H}[\rho]$ are unitarily equivalent if no term is omitted in $\bar{H}[\rho]$.}
\begin{align}
\bar{H}[\rho]
    \equiv& \,\,   {\bar h}^{(0)}[\rho] + {\bar h}^{(1)}[\rho] + {\bar h}^{(2)}[\rho]  \notag  \\
    \equiv& \,\,  {\bar h}^{(0)}[\rho] \notag  \\
    & \,\, +\frac{1}{(1!)^2} \sum_{\substack{a_1\\b_1}} {\bar h}^{a_1}_{b_1}[\rho] \, A^{a_1}_{b_1} \notag \\
    & \,\, +\frac{1}{(2!)^2} \sum_{\substack{a_1a_2\\b_1b_2}} {\bar h}^{a_1a_2}_{b_1b_2}[\rho] \, A^{a_1a_2}_{b_1b_2}  \, , \label{finalapproxH}
\end{align}
where the maximum-rank has been reduced to $k=2$ and where the pure number ${\bar h}^{(0)}[\rho]$ has now appeared~\cite{Frosini:2021tuj}. 

The set of mode-$2k$ tensors ($k\leq2$) defining $\bar{H}[\rho]$ depend on the system through the symmetry-invariant one-body density matrix $\rho$ of an auxiliary symmetry-conserving many-body state. Interestingly, there exists a large flexibility regarding the nature of that auxiliary state that can typically be different from the actual solution of the Schr\"odinger's equation associated to $H$ (or $\bar{H}[\rho]$) one is looking for or from the reference state employed to expand it~\cite{Frosini:2021tuj}. Using the density matrix extracted from a simple spherical Hartree-Fock-Bogoliubov calculation provides a good enough accuracy in practice.  For simplicity, the functional dependence of $\bar{H}[\rho]$ on $\rho$ is omitted below. Of course, if the rank-reduction approximation can be avoided, $\bar{H}$ must be replaced by $H$ everywhere below.

\subsubsection{PGCM-PT expansion method}

The recent stepping of ab initio calculations into the realm of mid-mass nuclei has essentially been due to the development and implementation of so-called {\it expansion many-body methods}~\cite{Hergert20,Tichai20,Frosini:2021fjf}. Generically, these methods rely on a partitioning of the Hamiltonian
\begin{align}
\bar{H} &=  \bar{H}_0 + \bar{H}_1 \,  \label{partitioninginitial}
\end{align}
chosen such that (at least) one appropriate eigenstate $| \Theta^{\sigma}_{\mu} \rangle$ of $\bar{H}_0$ is known, i.e. 
\begin{align}
\bar{H}_0 | \Theta^{\sigma}_{\mu} \rangle &=  E^{\tilde{\sigma}(0)}_{\mu} \, | \Theta^{\sigma}_{\mu} \rangle \, . \label{eignvaluerefstate}
\end{align}
The above partitioning of $\bar{H}$ materializes into the introduction of  two projectors in direct sum on ${\cal H}_{\text{A}}$
\begin{subequations}
\label{projectorsFockspace}
\begin{align}
{\cal P}^{\tilde{\sigma}}_{\mu} &\equiv  \sum_{K} | \Theta^{\tilde{\sigma}K}_{\mu} \rangle \langle \Theta^{\tilde{\sigma}K}_{\mu} | \, , \label{projectorsFockspace1} \\
{\cal Q}^{\tilde{\sigma}}_{\mu} &\equiv  1 - {\cal P}^{\tilde{\sigma}}_{\mu} \, . \label{projectorsFockspace2}
\end{align}
\end{subequations}
The operator ${\cal P}^{\tilde{\sigma}}_{\mu}$ projects on the eigen subspace of $\bar{H}_0$ spanned by the  {\it unperturbed state} $| \Theta^{\sigma}_{\mu} \rangle$ along with the degenerate states obtained via symmetry transformations, i.e. belonging to the same irreducible representation (IRREP) of the symmetry group of $\bar{H}$. This constitutes the so-called ${\cal P}$ space. The operator ${\cal Q}^{\tilde{\sigma}}_{\mu}$ projects onto the complementary orthogonal subspace, the so-called ${\cal Q}$ space.

Given the unperturbed state, expansion methods aim at finding an efficient way to connect it to a target eigenstate $| \Psi^{\sigma}_{\mu} \rangle$ of $\bar{H}$. This connection is formally achieved via the so-called {\it wave operator} $\Omega$, i.e.
\begin{align}
| \Psi^{\sigma}_{\mu} \rangle &\equiv \Omega | \Theta^{\sigma}_{\mu} \rangle \, , \label{waveoperator}
\end{align}
which is {\it state specific}\footnote{The state-specific character of the expansion means that each eigenstate is expanded independently from the others and that the wave operator carries a dependence on the state expanded, even though this dependency is left implicit in Eq.~\eqref{waveoperator} for simplicity.} and carries the complete effect of the residual interaction $\bar{H}_1$ in ${\cal Q}$.

In order to access all, i.e. doubly closed-shell, singly open-shell and doubly open-shell, nuclei in a controlled and polynomial fashion, the expansion method must rely on an unperturbed state $| \Theta^{\sigma}_{\mu} \rangle$ that (a) overcomes the potential occurrence of (near) degenerate elementary excitations, (b) carries good symmetry quantum numbers and (c) solves Eq.~\eqref{eignvaluerefstate} at low numerical cost. While the last point imposes strong constraints on the affordable complexity of the unperturbed state, items (a) and (b) push in the opposite direction. Indeed, the first two points require that so-called {\it strong/static} correlations arising in open-shell nuclei are already incorporated into the unperturbed state in order for the subsequent expansion to be well defined and limited to capturing gentler {\it weak/dynamical} correlations.  

Recently, this tension could be resolved via the formulation~\cite{Frosini:2021fjf} and the implementation~\cite{Frosini:2021ddm} of a novel expansion method coined as the {\it projected generator coordinate method perturbation theory} (PGCM-PT). The PGCM-PT formalism designs a perturbative expansion around a {\it multi-reference symmetry-invariant} unperturbed state obtained through the PGCM and thus reading as
\begin{align}
\ket{\mathrm \Theta^{\sigma}_{\mu} }  
		&= \sum_q f^{\tilde{\sigma}}_{\mu}(q) P^{\tilde{\sigma}}_{M0} | \Phi (q) \rangle \, . \label{PGCMstate}
\end{align}
In Eq.~\eqref{PGCMstate}, $\text{B}_{q} \equiv \{ \ket{\Phi(q)}; q \in \text{set} \}$ denotes a set of non-orthogonal Bogoliubov product states differing by the value of the collective deformation parameter $q$. These reference states are typically obtained in a first step by repeatedly solving Hartree-Fock-Boboliubov (HFB) mean-field equations\footnote{When HFB equations reduce to spherical HF ones in doubly closed-shell nuclei, employing $\bar{H}$ is strictly equivalent to using $H$ in the mean-field step.} with a Lagrange term associated with a constraining operator\footnote{The generic operator $Q$ can embody several constraining operators such that the collective coordinate $q$ may in fact be multi dimensional.} $Q$ such that the solution satisfies
\begin{align}
\langle \Phi(q) | Q | \Phi(q) \rangle &= q \, . \label{constraint}
\end{align} 
The  operator $Q$ is typically chosen such that $| \Phi(q) \rangle$ breaks symmetry(ies) of the Hamiltonian as soon as $q\neq 0$. Because physical states must carry good symmetry quantum numbers \((\tilde{\sigma},M)\), the projector $P^{\tilde{\sigma}}_{M0}$ ensures that it is the case of $\ket{\mathrm \Theta^{\sigma}_{\mu} }$; see Ref.~\cite{Frosini:2021fjf} for details regarding the symmetry group and the symmetry projector at play. 

The unknown coefficients $\{ f^{\tilde{\sigma}}_{\mu}(q); q \in \text{set} \}$ entering $\ket{\mathrm \Theta^{\sigma}_{\mu} }$ are determined via the application of Ritz' variational principle within the PGCM subspace ${\cal H}^{\text{PGCM}}_{\text{A}}$. The latter is spanned by the set $\text{PB}_{q\tilde{\sigma}} \equiv \{ P^{\tilde{\sigma}}_{M0}\ket{\Phi(q)}; q \in \text{set} \}$ of non-orthogonal {\it projected} HFB states\footnote{While the span of ${\cal H}^{\text{PGCM}}_{\text{A}}$ is  limited by the generic form of the PGCM ansatz in Eq.~\eqref{PGCMstate}, the nature and number of collective coordinates $q$ is not circumscribed such that the richness of the PGCM ansatz can vary quite significantly depending of the choices made by the practitioner. Eventually, the possible span of the MR-EDF ansatz remains unclear.}. This leads to solving Hill-Wheeler-Griffin's equation~\cite{RiSc80}
\begin{align}
\sum_{q} H^{\tilde{\sigma}}_{pq} \, f^{\tilde{\sigma}}_{\mu}(q) &=  {\cal E}^{\tilde{\sigma}}_{\mu} \sum_{q}  N^{\tilde{\sigma}}_{pq} \, f^{\tilde{\sigma}}_{\mu}(q) \, , \label{HWG_equation}
\end{align}
which actually delivers a {\it set} of PGCM states\footnote{The diagonalization is performed separately for each value of $\tilde{\sigma}$, i.e. within each IRREP of $\text{G}_{H}$, such that one generates ground and excited states for each possible combination of the symmetry quantum numbers.} with associated PGCM energies 
\begin{align}
{\cal E}^{\tilde{\sigma}}_{\mu} &= \frac{\langle \Theta^{\sigma}_{\mu} | \bar{H} | \Theta^{\sigma}_{\mu} \rangle}{\langle \Theta^{\sigma}_{\mu} | \Theta^{\sigma}_{\mu} \rangle}  \, . \label{PGCMenergy}
\end{align}
Each PGCM state coming out of Eq.~\ref{HWG_equation} can play the role of the unperturbed state for a target eigenstate to be reached according to Eq.~\ref{waveoperator}.

The PGCM ansatz is characterized by its capacity to efficiently capture strong static correlations at play in open-shell nuclei from a low-dimensional, i.e. from several tens to a few hundreds, configuration mixing at the price of dealing with non-orthogonal vectors in the computation of the Hamiltonian ($H^{\tilde{\sigma}}_{pq}$) and norm ($N^{\tilde{\sigma}}_{pq}$) kernels~\cite{RiSc80}. Such a PGCM unperturbed state renders the method universal, i.e. automatically well-behaved for doubly closed-shell, singly open-shell and doubly open-shell nuclei. In the case of closed-shell systems only displaying weak dynamical correlations, reducing the PGCM ansatz to a single symmetry-conserving Slater determinant is safe and delivers traditional single-reference expansion methods. 

Selecting one of the PGCM states coming out of Eq.~\ref{HWG_equation} as the unperturbed state $\ket{\mathrm \Theta^{\sigma}_{\mu} }$, the associated {\cal P} and {\cal Q} spaces are set. Next, an appropriate ab initio method must expand and truncate the wave operator $\Omega$ to capture the remaining (i.e. dynamical) correlations originating from ${\cal Q}$ to access the target state $\ket{\mathrm \Psi^{\sigma}_{\mu} }$. In Ref.~\cite{Frosini:2021fjf}, the expansion was realized designed perturbatively, thus introducing the state-specific PGCM-PT wave operator $\Omega^{\text{PGCM-PT}}$. Truncating the expansion to second order provides the PGCM-PT(2) approximation to an exact eigenvalue of $H$ through\footnote{The asymmetric form of the exact energy employed in Eq.~\eqref{ref_E_PGCM} is said to be {\it projective} because it originates from left-projecting Eq.~\eqref{schroedevolstate} onto the unperturbed bra $\langle \Theta^{\sigma}_{\mu} |$. This is indeed the form from which, e.g., perturbation theories or basic coupled cluster theories are formulated.}
\begin{align}
E^{\tilde{\sigma}}_{\mu} &= \frac{\langle \Theta^{\sigma}_{\mu} | \bar{H} | \Psi^{\sigma}_{\mu} \rangle}{\langle \Theta^{\sigma}_{\mu} | \Psi^{\sigma}_{\mu} \rangle} \nonumber \\
&= \sum_{k=0}^{\infty} E^{\tilde{\sigma}(k)}_{\mu} \nonumber \\
&\approx {\cal E}^{\tilde{\sigma}}_{\mu} + E^{\tilde{\sigma}(2)}_{\mu} , \label{ref_E_PGCM}
\end{align}
where the PGCM energy [Eq.~\eqref{PGCMenergy}] of the reference state satisfies ${\cal E}^{\tilde{\sigma}}_{\mu}=E^{\tilde{\sigma}(0)}_{\mu}+E^{\tilde{\sigma}(1)}_{\mu}$ and where the calculation of the second-order correction $E^{\tilde{\sigma}(2)}_{\mu}$ is detailed in Ref.~\cite{Frosini:2021fjf}.

\subsection{EDF many-body formalism}
\label{MREDF}

\subsubsection{Prolegomenon}

Because the nuclear EDF formalism currently relies on an empirical formulation, it still resists a truly rigorous definition and is not systematically improvable. Several interpretations have been provided throughout the years, most of which are at best incomplete and do not often correspond to what EDF calculations {\it actually} encompass in practice. Because the goal of this work is not to dispute fallacious interpretations, the formulation below pragmatically focuses on what current EDF calculations are. 

Knowing that one should always start from the most general formulation out of which more limited versions derive as particular cases, the nuclear EDF formalism is presently understood as what is traditionally denoted as the {\it multi-reference} EDF (MR-EDF) method from which the single-reference level (SR-EDF) can indeed be deduced as a particular subcase.

\subsubsection{Theoretical scheme}

The EDF approach~\cite{10.1088/2053-2563/aae0ed} also employs (i) A non-relativistic point-like nucleons as degrees of freedom. It however postulates that the A-body dynamics of these nucleons is restricted to the PGCM subspace ${\cal H}^{\text{PGCM}}_{\text{A}}$. Thus, (ii) Schr\"{o}dinger's equation delivering many-body observables is projected onto the subspace ${\cal H}^{\text{PGCM}}_{\text{A}}$ of the A-body Hilbert space  ${\cal H}_{\text{A}}$ obtained via the (anti-symmetrized) tensor product of A one-body Hilbert spaces ${\cal H}_{\text{1}}$ associated with A non-relativistic point-like nucleons\footnote{As opposed to, e.g., valence-space shell model techniques, the EDF method does not proceed through a partitioning of the one-body Hilbert space ${\cal H}_{\text{1}}$ via the definition of a so-called valence/active space but directly restricts the resulting A-body Hilbert space ${\cal H}_{\text{A}}$.}. Correspondingly, the EDF method (iii) postulates the existence of an effective Hamiltonian $H_{\text{EDF}}$ that can produce, within ${\cal H}^{\text{PGCM}}_{\text{A}}$, the same low-energy observables as the ab initio scheme formulated in  ${\cal H}_{\text{A}}$, e.g. low-lying A-body eigenenergies can in principle satisfy
\begin{align}
E^{\tilde{\sigma}}_{\mu} &= \frac{\langle \Theta^{\sigma}_{\mu} | H_{\text{EDF}} | \Theta^{\sigma}_{\mu} \rangle}{\langle \Theta^{\sigma}_{\mu} | \Theta^{\sigma}_{\mu} \rangle} \, . \label{E_MREDF}
\end{align}

\begin{figure*}
    \centering
    \includegraphics[width=\textwidth]{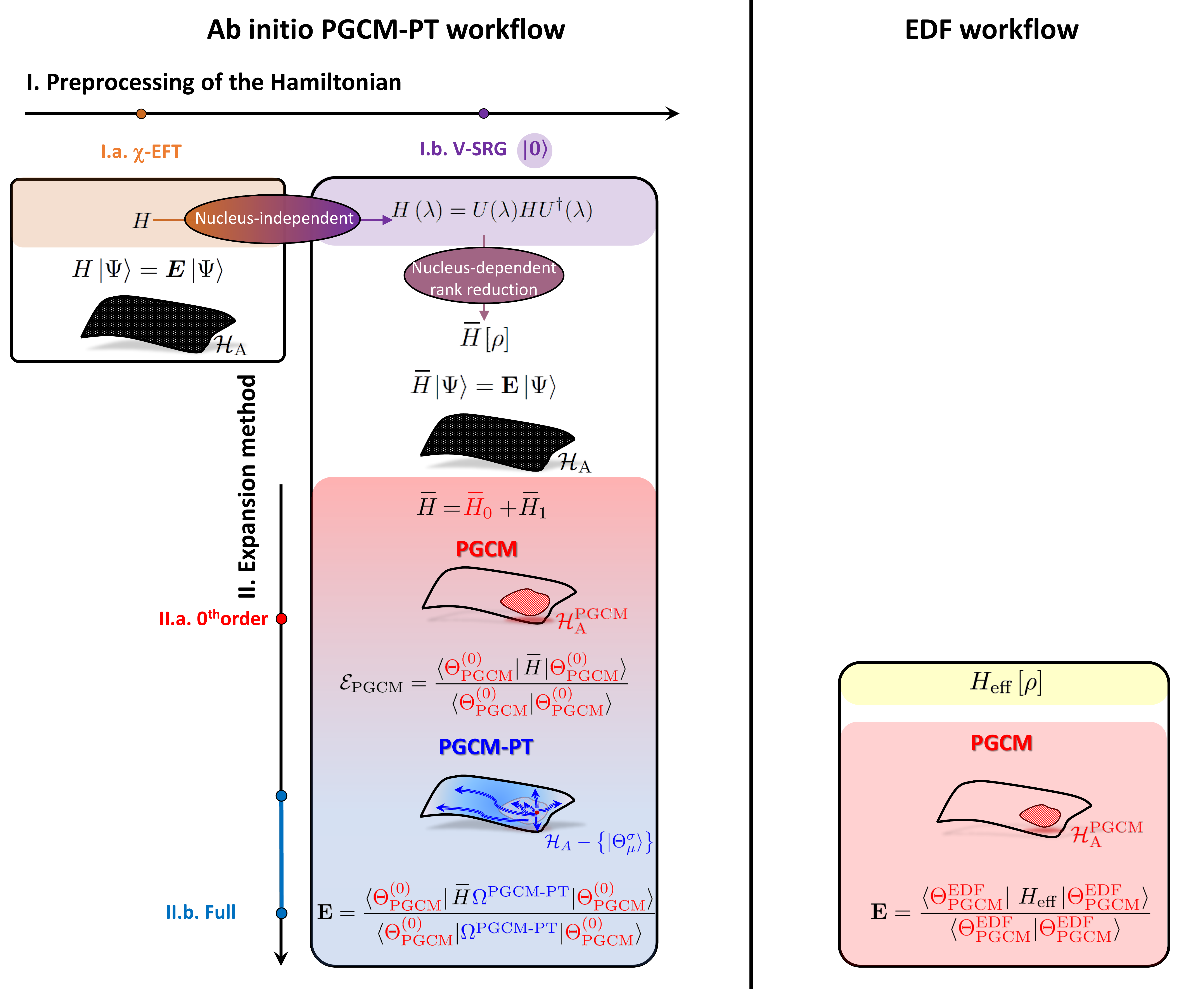}
    \caption{Schematic workflows of the two theoretical schemes. Left: the PGCM-PT ab initio many-body expansion method. Right: the (so far empirical) nuclear energy density functional many-body method.}
    \label{fig:workflow}
\end{figure*}

\subsubsection{Formal difficulty}
\label{spuriosities}

Until now, the EDF method has relied on empirically-built, nucleus-dependent, effective Hamiltonians $H_{\text{EDF}}$. The nucleus dependency, crucial from the phenomenological standpoint, has been typically expressed in terms of the local one-nucleon density distribution. Originally formulated at the SR-EDF level, i.e. whenever the PGCM state reduces to a single HFB product state $| \Phi(q) \rangle$, the nucleon density was naturally computed out of this many-body state. However, the freedom left to extend this empirical ansatz to the MR-EDF case, i.e. to the PGCM ansatz\footnote{For several proposed options, writing the EDF energy $E^{\tilde{\sigma}}_{\mu}$ as in Eq.~\eqref{E_MREDF}, i.e. as the expectation of an operator in the PGCM state is not even valid~\cite{Duguet:2013dga}.}, has generated intense discussions and shown to lead to problematic pathologies of various sorts depending on the choice made; see Ref.~\cite{Sheikh:2019qdz} and references therein for a recent summary of the issue and of the various attempts to address it. 

\subsection{Discussion}

Before coming to a numerical comparison, let us briefly contrast the formal ingredients entering the two theoretical schemes whose workflows are schematically compared in Fig.~\ref{fig:workflow}. 

\begin{figure*}
    \centering
    \includegraphics[width=0.7\textwidth]{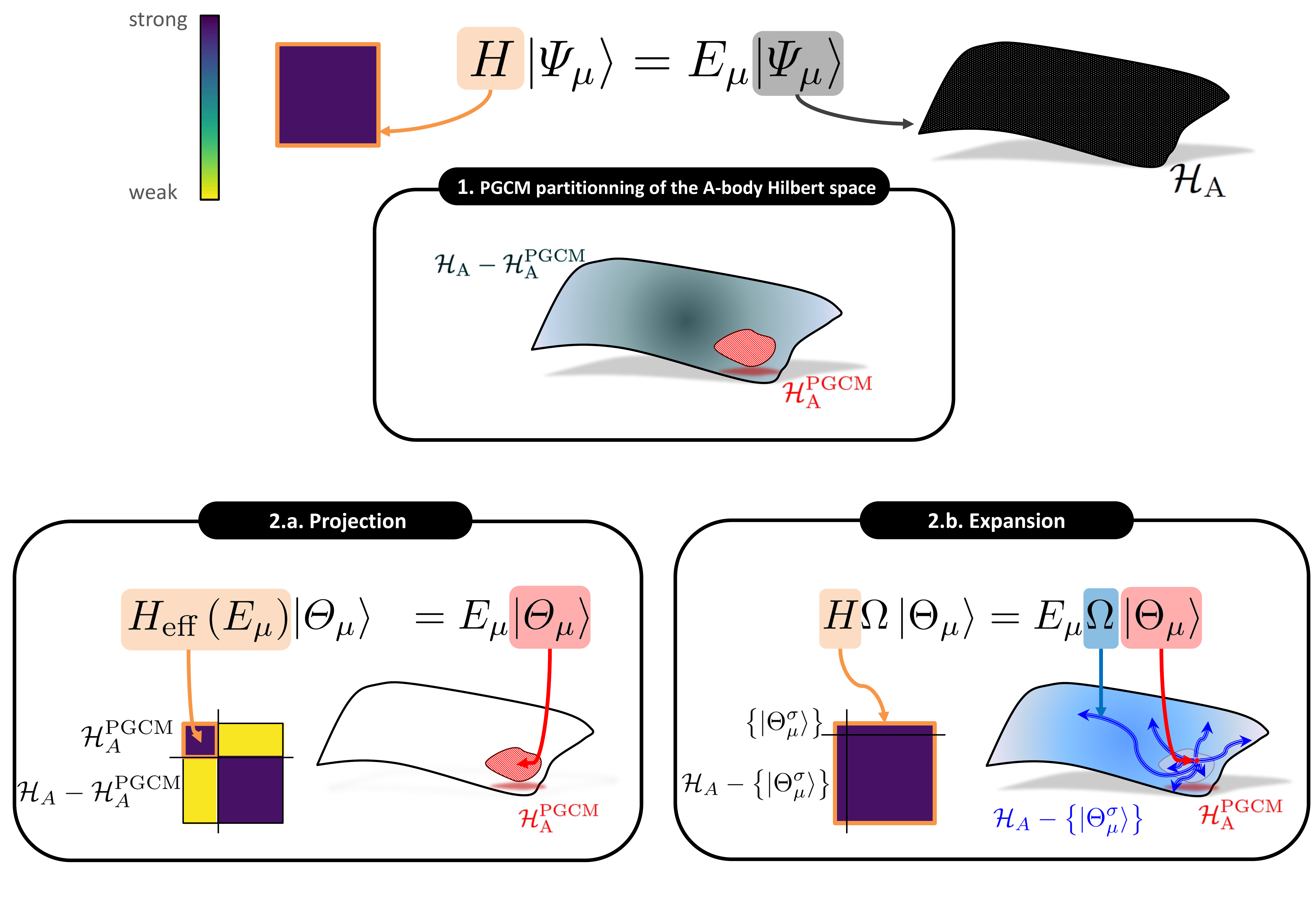}
    \caption{Many-body strategies to solve A-body Schr\"{o}dinger's equation via the PGCM partitioning of the A-body Hilbert space ${\cal H}_{\text{A}}$ into two subspaces ${\cal H}^{\text{PGCM}}_{\text{A}}$ and ${\cal H}_{\text{A}} - {\cal H}^{\text{PGCM}}_{\text{A}}$. In option 2.a., A-body Schr\"{o}dinger's equation is projected onto ${\cal H}_{\text{A}}$ to deliver the targeted eigenvalue out of an effective energy-dependent Hamiltonian. In option 2.b., the targeted eigenstate in ${\cal H}_{\text{A}}$ is expanded in terms of an eigenstate of the unperturbed Hamiltonian $\bar{H}_0$ belonging to ${\cal H}^{\text{PGCM}}_{\text{A}}$ and of a wave operator $\Omega$ taking explicit care of the correlations induced by the residual interaction $\bar{H}_1$.}
    \label{fig:partitioning_vs_projection}
\end{figure*}

For the first time, PGCM-PT provides a formally exact ab initio method whose starting point, i.e. the unperturbed state on which the many-body expansion is based, is the PGCM ansatz situated at the heart of the EDF method. According to the schematic representation displayed in Fig.~\ref{fig:partitioning_vs_projection}, PGCM-PT follows strategy 2.b., i.e. while static correlations are captured within ${\cal H}^{\text{PGCM}}_{\text{A}}$, dynamical correlations are explicitly computed via the perturbative expansion of the wave operator $\Omega$ driven by $\bar{H}_1$. 

The EDF method relies on the same partitioning of ${\cal H}_{\text{A}}$, which is eventually crucial to connect the two theoretical schemes. While static correlations are similarly captured within ${\cal H}^{\text{PGCM}}_{\text{A}}$, dynamical correlations are however now meant to be entirely compensated for via the use of $H_{\text{EDF}}$\footnote{In principle, elementary excitations can be incorporated into the PGCM ansatz itself as an extra degree of freedom to capture dynamical correlations. However, doing so beyond a few dominant configurations would contradict the key point to keep the associated diagonalization (Eq.~\ref{HWG_equation}) low dimensional, which is mandatory to maintain the low numerical scaling of the EDF method (see below). In other words, the bulk of dynamical correlations are not meant to be grasped through the PGCM state.}. As such, the EDF method follows, at least in principle, strategy 2.a. in Fig.~\ref{fig:partitioning_vs_projection}, i.e. it proceeds through a projection on ${\cal H}^{\text{PGCM}}_{\text{A}}$. In the current empirical formulation of the EDF method, this projection is only implicit and relies on the hypothesis, never explicitly demonstrated so far, that dynamical correlations vary only smoothly with particle number and can be universally parameterized and captured through a simplistic density dependence of the effective Hamiltonian $H_{\text{EDF}}$ acting within ${\cal H}^{\text{PGCM}}_{\text{A}}$. 

The specific parametrization of $H_{\text{EDF}}$, and/or of the EDF energy $E^{\tilde{\sigma}}_{\mu}$, in terms of the one-body (local) density has become the central point of discussion in the last twenty years due to the (dubious) temptation to root the nuclear EDF method into the Hohenberg-Kohn theorem of electronic density functional theory (DFT). While it might eventually be the most practical way to parameterize the nucleus dependency of $H_{\text{EDF}}$, the formulation in terms of the nucleon density distribution is not the most crucial characteristic of the EDF method (or its successor of present interest). Eventually, the key promise of the EDF method is to access essentially exact results at the low numerical cost associated with the restriction of the A-body dynamics to the PGCM subspace of ${\cal H}_{\text{A}}$ and authorize repeated calculations over the entire nuclear chart, i.e. of thousands of nuclei, on a short time scale.

\begin{table*}[h]
    \centering
    \begin{tabular}{|l|c|c|c|c|}
    \hline
       Method  & HFB & PGCM &  PGCM-PT(2) & FCI \\
           \hline
Runtime & \(O(n_{\text{dim}}^4)\) & \(O(n_{\text{proj}} 
n_\text{gcm}^2 n_{\text{dim}}^4)\)  &  \(O(n_{\text{proj}} n_\text{gcm}^2 n_{\text{dim}}^8)\) & \(O(n_{\text{dim}}^{\text{A}})\) \\
       Storage & \(O(n_{\text{dim}}^4)\) & \(O(n_{\text{dim}}^4)\)   & \(O(n_\text{gcm}^2 n_{\text{dim}}^8)\) & \(O(n_{\text{dim}}^{\text{A}})\) \\
           \hline
    \end{tabular}
    \caption{Runtime complexity and storage requirements for various resolution methods of the many-body problem. While $n_{\text{dim}}$ denotes the dimension of the (truncated) one-body Hilbert space ${\cal H}_{\text{1}}$, \(n_{\text{proj}}\) represents the number of angles used to discretize the symmetry projector $P^{\tilde{\sigma}}_{M0}$ and \(n_\text{gcm}\) the number of states used in the mixing over $q$.}
    \label{tab:complexity}
\end{table*}

Table~\ref{tab:complexity} provides the naive\footnote{The ``naive" character of the scaling first relates to the fact that the exploitation of symmetries can typically reduce the effective cost of any given method. Moreover, the evaluation of the EDF cost is naive given that the production of $H_{\text{EDF}}$, which is indeed null in the current empirical formulation of the EDF method, is omitted.} numerical scaling of the two theoretical schemes as a function of the basis size dimension $n_{\text{dim}}$ of the (truncated) one-body Hilbert space ${\cal H}_{\text{1}}$, which constitutes the main variable driving that scaling. While the exact solution, i.e. full configuration interaction (FCI), faces the exponential scaling of ${\cal H}_{\text{A}}$, PGCM-PT(2) displays a polynomial cost dominated by the second-order correction that scales as $n_{\text{dim}}^8$. Contrarily, the PGCM step, which constitutes the full cost of the EDF method, displays a much milder $n_{\text{dim}}^4$ scaling\footnote{The key advantage of the PGCM method is that the prefactor associated with \(n_{\text{proj}}\), the number of angles used to discretize the symmetry projector $P^{\tilde{\sigma}}_{M0}$ and with \(n_\text{gcm}\), the number of states used in the mixing over $q$, although notable, displays a very mild scaling with nucleon number A and thus with $n_{\text{dim}}$.}. Bypassing the explicit resummation of dynamical correlations and its associated computational cost is what makes the EDF method appealing. The possibility to limit many-body calculations to the EDF scaling in a controlled fashion, i.e. in a way that is rooted into the ab initio paradigm, is at the heart of the discussion below. 

\section{Numerical comparison}
\label{sec:2}

Having established a formal connection between the two theoretical schemes, the objective is now to compare the associated results for an illustrative example, i.e. the ground-state binding energy and rotational band of the doubly open-shell $^{20}$Ne nucleus, and draw some useful conclusions\footnote{While the present numerical illustration employs a specific chiral Hamiltonian and a specific EDF parametrization, the outcome and conclusions are generally valid.}. 

\begin{figure}
    \centering
    \includegraphics[width=0.5\textwidth]{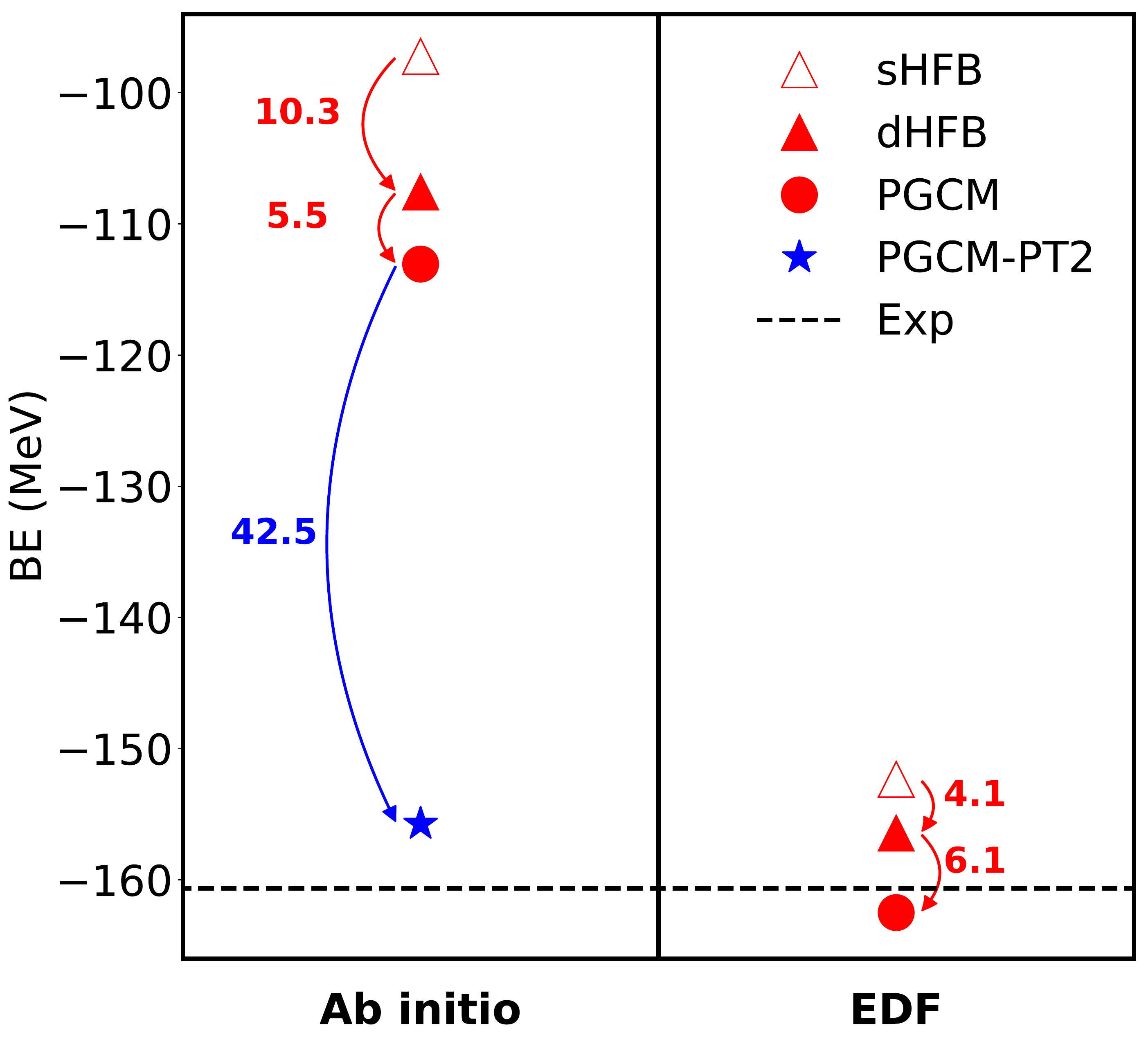}
    \caption{(color online) Successive contributions to the absolute ($J^{\pi}=0^+$) ground-state energy of $^{20}$Ne obtained from (left) PGCM-PT(2) and (right) EDF calculations. See text for details.}
    \label{fig:ebran_plots}
\end{figure}

The PGCM-PT calculation employs a VSRG-evolved chiral N${}^{3}$LO nucleon-nucleon interaction with \(\lambda_{\text{vsrg}}=1.8\text{ fm}^{-1}\), supplemented with an N${}^{2}$LO three-nucleon interaction with cutoff $\Lambda=2.0\,\mathrm{fm}^{-1}$ whose low-energy constants are adjusted to $A=3,4$ observables, as described in Refs.~\cite{PhysRevC.83.031301,Nogga:2004il}. The Hamiltonian $H$ is turned into an in-medium two-body operator $\bar{H}[\rho]$ via the rank reduction method~\cite{Frosini:2021tuj} discussed in Sec.~\ref{rankred}. The EDF calculation utilizes the relativistic point coupling parametrization DD-PC1~\cite{PhysRevC.78.034318}. Both calculations employ a spherical harmonic oscillator (HO) one-body basis characterized by the parameter $\hbar \omega=20\mathrm{\,MeV}$ and $e_{\mathrm{max}}=6$. In both cases, the PGCM state at play mixes axially-deformed HFB states constrained to the axial quadrupole moment ($q\equiv q_{20}$) $\beta_2 \in [0.3,0.8]$ and further projected on good neutron $N$ and proton $Z$ numbers as well as on good total angular momentum $J$.

Figure~\ref{fig:ebran_plots} displays the successive contributions to $^{20}$Ne ground-state energy, i.e. (a) spherical HFB (sHFB), (b) deformed HFB (dHFB), (c) PGCM and, for the ab initio calculation, adding the (d) PGCM-PT(2) correction. The striking observation is that the energies generated through the first three steps are very different in the two calculations. While the PGCM step is final in the EDF calculation, and thus close to the experimental value, the corresponding ab initio result is about 45\,MeV unbound. Thankfully, this deficit is consistently compensated for by the explicit inclusion of dynamical correlations via the second-order PGCM-PT(2) correction.  While both calculations are eventually satisfactory, they reach the end result in an apparently very different fashion in spite of employing the same technique to grasp static correlations via the PGCM. 

This apparent mismatch and the key role played by explicit dynamical correlations in the ab initio calculation seem to forbid a practical connection between the two theoretical schemes. Under closer inspection, one observes that the scales at play in steps (b) (sHFB$\longrightarrow$dHFB) and (c) (dHFB$\longrightarrow$PGCM) taking care of static correlations are, if not identical, actually consistent in both calculations. Eventually, the difference mostly correlates with a shift up of the ab initio starting point (sHFB) that is essentially compensated for by the PGCM-PT(2) correction. This examination indicates that there might be a way to relate the two schemes in a more transparent and explicit fashion. 

\begin{figure}
    \centering
    \includegraphics[width=0.45\textwidth]{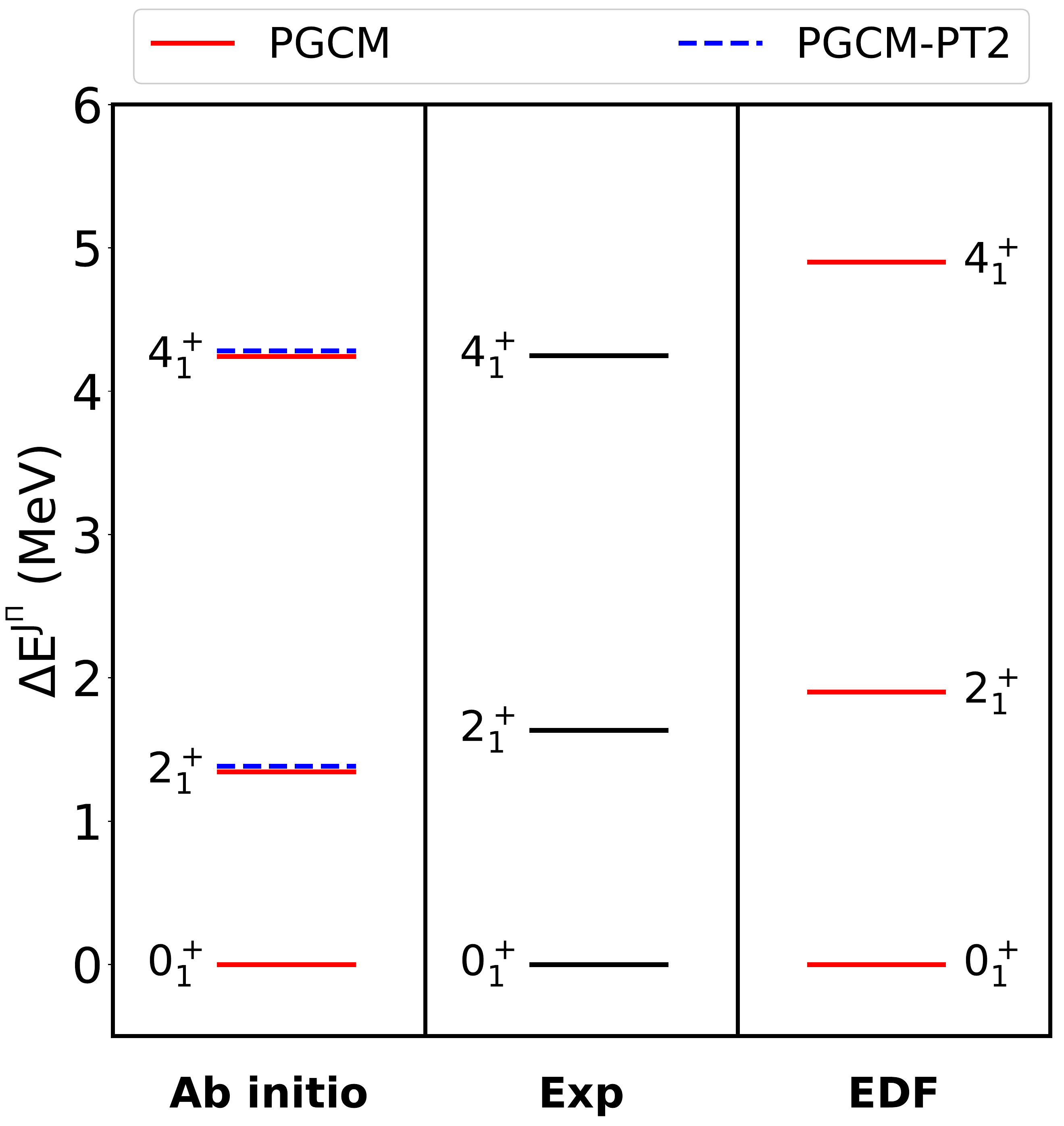}
    \caption{(color online) Low-lying part of $^{20}$Ne ground-state rotational band. Left: PGCM-PT(2) calculation. Middle: experimental data. Right: EDF calculation. See text for details.}
    \label{fig:ebran_plots_spectrum}
\end{figure}

Before coming to this key point, the analysis is extended to the low-lying part of \nucl{Ne}{20} ground-state rotational band in Fig.~\ref{fig:ebran_plots_spectrum}. Quite strikingly, the ab initio PGCM spectrum is in excellent agreement with experiment\footnote{These results are also in good agreement with FCI calculations~\cite{Frosini:2021ddm}.}. Consistently, the large PGCM-PT(2) correction ($\sim45$\,MeV) to each absolute energy cancels out to a high degree and leave the spectrum essentially unchanged, making ab initio PGCM calculations already well suited to describe low-lying rotational bands. As seen from the right panel, the EDF spectrum is slightly too spread out. While the degree of dilatation depends on the specific parameterization used and happens to be rather moderate with DD-PC1, this dilatation is a generic feature of rotational bands obtained in PGCM calculations performed within the EDF context whenever based on the sole axial quadrupole shape degree of freedom. 

\section{Pre-processing of the Hamiltonian}
\label{sec:3}

When computing total energies, as in the example displayed in Fig.~\ref{fig:ebran_plots}, it is not surprising that the first three steps of ab initio and EDF calculations lead to different results. Indeed, although based on the same (i.e. PGCM) wave-function ansatzes, the two calculations differ by the very nature of the employed Hamiltonians, i.e. while $H$ is defined over ${\cal H}_{\text{A}}$, $H_{\text{EDF}}$ effectively operates in a subspace of it. Thus, the key to a consistent connection between the two theoretical schemes relies on a meaningful link between the two Hamiltonians at play. A formally transparent approach to connect the Hamiltonian $H$ acting on ${\cal H}_{\text{A}}$ to an effective Hamiltonian whose action is restricted to the subspace ${\cal H}^{\text{PGCM}}_{\text{A}}$ is provided by (non-unitary) subspace projection techniques~\cite{feshbach58a,feshbach62a}. This procedure typically produces an energy- and system-dependent A-body effective Hamiltonian that delivers the same low-energy eigenvalues as $H$ while acting on the targeted subspace. In order to avoid the complexity of dealing with an energy-dependent Hamiltonian, the use of multi-reference unitary in-medium similarity renormalization group (MR-IMSRG) transformations~\cite{Tsukiyama:2010rj,Hergert16} of the initial Hamiltonian is presently advocated to effectively decouple ${\cal H}^{\text{PGCM}}_{\text{A}}$ from the complementary subspace of ${\cal H}_{\text{A}}$.

\subsection{MR-IMSRG transformations}
\label{sec:mrimsrg}

As for an actual presentation of MR-IMSRG transformation techniques, the reader is referred to Refs.~\cite{Hergert16,Hergert:2016iju,Hergert:2016etg}. For the present purpose, it is sufficient to stipulate the existence of a continuous manifold of {\it nucleus-dependent}\footnote{Rigorously speaking, the transformation is {\it reference-state-dependent}. Considering a reference state in the appropriate A-body Hilbert space for each system under study, e.g. the PGCM state $| \Theta^{\sigma}_{\mu} \rangle$ in the present case, the MR-IMSRG transformation ${\cal U}(s)$ indeed carries an implicit nucleus dependency.} unitary transformations ${\cal U}(s)$ parameterized by a momentum scale $s$ and tailored such that the PGCM reference state becomes largely decoupled\footnote{In closed-shell nuclei, the simpler single-reference IMSRG (SR-IMSRG) method relying on a rotationally invariant Slater determinant reference state can be meaningfully applied. In this case, taking $s\rightarrow\infty$ leads to a complete decoupling of the reference state from the rest of the Hilbert space, effectively resumming all dynamical correlations into the pre-processed Hamiltonian $H(\infty)$ such that no correlations need to be further added. In the more general case of present interest, the PGCM state cannot be fully decoupled from the rest of the Hilbert space. Thus, while dynamical correlations are largely resummed via the MR-IMSRG pre-processing as shown below, an additional step, e.g. PGCM-PT, is always needed to grasp the remaining correlations originating from the complementary ${\cal Q}$ space.} from the orthogonal subspace of ${\cal H}_{\text{A}}$, the ${\cal Q}$ space, via the transformed Hamiltonian $\bar{H}(s) \equiv {\cal U}(s) \bar{H} {\cal U}^{\dagger}(s)$ when $s\rightarrow\infty$. 

\begin{figure*}
    \centering
    \includegraphics[width=\textwidth]{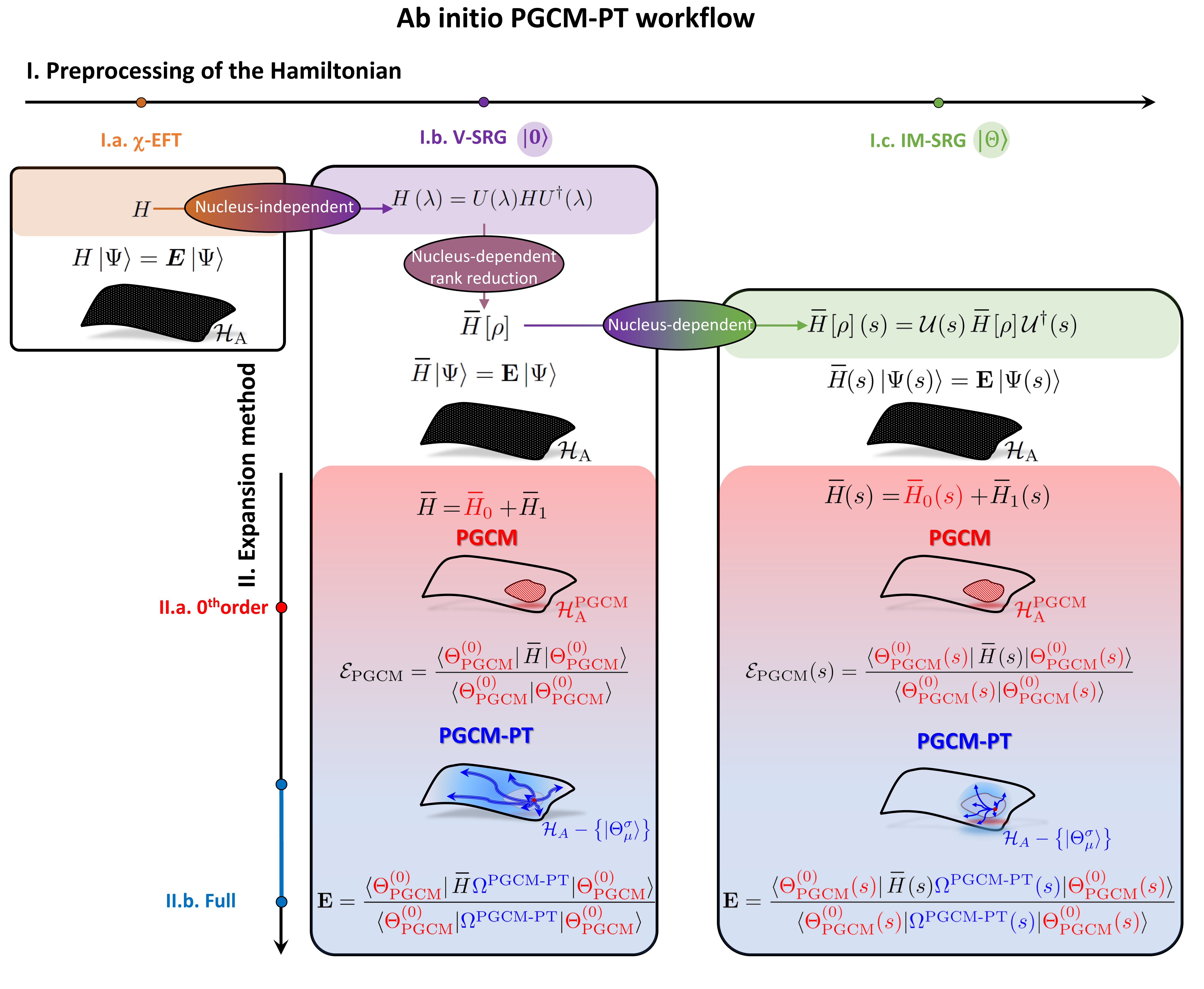}
    \caption{Extended ab initio PGCM-PT workflow including the MR-IMSRG pre-processing of the Hamiltonian.}
    \label{fig:workflow2}
\end{figure*}

Applying the unitary transformation to Eq.~\eqref{schroedevolstate} formally leads to
\begin{align}
\bar{H}(s) | \Psi^{\sigma}_{\mu}(s) \rangle &=  E^{\tilde{\sigma}}_{\mu} \, | \Psi^{\sigma}_{\mu}(s) \rangle \, , \label{transformedschroed}
\end{align}
with
\begin{align}
| \Psi^{\sigma}_{\mu}(s) \rangle &\equiv  {\cal U}(s) | \Psi^{\sigma}_{\mu} \rangle \nonumber \\
&\equiv \Omega^{\text{PGCM-PT}}(s) | \Theta^{\sigma}_{\mu}(s) \rangle \, , \label{transformedstate}
\end{align}
where the last line testifies that the PGCM state acquires an explicit $s$ dependence given that it is eventually re-computed from the transformed Hamiltonian $\bar{H}(s)$. As opposed to the (energy-dependent) Hamiltonian obtained via non-unitary subspace projection techniques, $\bar{H}(s)$ still acts on the complete A-body Hilbert space. However, it does so in such a way that the PGCM state becomes more and more decoupled from the orthogonal subspace as $s\rightarrow\infty$. The MR-IMSRG pre-processing of $\bar{H}$ adds one layer to the ab initio workflow as schematically illustrated in Fig.~\ref{fig:workflow2} where the intensity of the coupling between ${\cal P}$ and ${\cal Q}$ subspaces is reduced as $s$ increases.

While the full eigenvalue $E^{\tilde{\sigma}}_{\mu}$ remains invariant under the unitary transformation ${\cal U}(s)$ as stipulated by Eq.~\eqref{transformedschroed}, wave-functions and intermediate contributions to the total energy all vary with $s$. In fact, Eqs.~\eqref{transformedschroed} and~\eqref{transformedstate} indicate that correlations are shifted from the initial wave-function to the Hamiltonian through the transformation ${\cal U}(s)$ such that a reshuffling of correlations shall occur as a function of $s$ while maintaining the total energy unchanged. If the PGCM state could be entirely decoupled from the complementary subspace at the end of the flow, the identity $\Omega^{\text{PGCM-PT}}(\infty)=1$ would hold. As mentioned earlier, complete decoupling can be achieved in principle for SR-IMSRG transformations based on Slater determinant reference states, but typically not in the more general MR-IMSRG case.  

The MR-IMSRG represents the Hamiltonian and other observables in a basis of second-quantized operators $:A^{a_1a_2\ldots}_{b_1b_2\ldots}:$ in normal order with respect to the \emph{correlated} unperturbed state, i.e. the PGCM state $| \Theta^{\sigma}_{\mu} \rangle$ in the present case. This is done employing the generalized normal ordering and Wick's theorem developed by Mukherjee and Kutzelnigg \cite{Kutzelnigg:1997fk} and leads to 
\begin{align}
    \bar{H}(s) \equiv& \,\, {\cal U}(s)\bar{H}{\cal U}^{\dagger}(s)  \notag\\
        \equiv& \,\,  {\bar h}^{(0)}(s) + :{\bar h}^{(1)}(s): + :{\bar h}^{(2)}(s): +\ldots \notag  \\
    \equiv& \,\, {\bar h}^{(0)}(s) \notag \\
    & \,\,+ \frac{1}{(1!)^2} \sum_{\substack{a_1\\b_1}} {\bar h}^{a_1}_{b_1}(s) \, :A^{a_1}_{b_1}:\notag \\
    & \,\,+\frac{1}{(2!)^2} \sum_{\substack{a_1a_2\\b_1b_2}} {\bar h}^{a_1a_2}_{b_1b_2}(s) \, :A^{a_1a_2}_{b_1b_2}: \notag \\
    & \,\,+\frac{1}{(3!)^2}  \sum_{\substack{a_1a_2a_3\\b_1b_2b_3}} {\bar h}^{a_1a_2a_3}_{b_1b_2b_3}(s) \, :A^{a_1a_2a_3}_{b_1b_2b_3}: \notag \\
    & \,\,+\ldots \, ,  \label{IMsrgH}
\end{align}
where $:\ldots:$ denotes the normal ordering with respect to $| \Theta^{\sigma}_{\mu} \rangle$. In the process, the tensors defining the evolved Hamiltonian not only acquire a dependence on $s$ as explicitly indicated in Eq.~\eqref{IMsrgH} but also become functionals of the set of irreducible density matrices associated with $| \Theta^{\sigma}_{\mu} \rangle$
    \begin{subequations}\label{eq:irred_density}
    \begin{align}
        \lambda^{a_1}_{b_1} &\equiv \bra{\Theta^{\sigma}_{\mu}}A^{a_1}_{b_1}\ket{\Theta^{\sigma}_{\mu}}\,,\\
        \lambda^{a_1a_2}_{b_1b_2} &\equiv \bra{\Theta^{\sigma}_{\mu}}A^{a_1a_2}_{b_1b_2}\ket{\Theta^{\sigma}_{\mu}} - \mathcal{A}\left(\lambda^{a_1}_{b_1}\lambda^{a_2}_{b_2}\right)\,,\\
        \lambda^{a_1a_2a_3}_{b_1b_2b_3} &\equiv \bra{\Theta^{\sigma}_{\mu}}A^{a_1a_2a_3}_{b_1b_2b_3}\ket{\Theta^{\sigma}_{\mu}} - \mathcal{A}\left(\lambda^{a_1}_{b_1}\lambda^{a_2}_{b_2}\right)\notag\\
        &\qquad - \mathcal{A}\left(\lambda^{a_1a_2}_{b_1b_2}\lambda^{a_3}_{b_3}\lambda^{a_3}_{b_3}\right)\,,\\
        &\vdots\notag
    \end{align}
    \end{subequations}
    where the antisymmetrizer $\mathcal{A}$ generates all unique index permutations of its arguments that are required to ensure overall antisymmetry; i.e. 
    \begin{equation}
    {\bar h}^{a_1\ldots}_{b_1\ldots}(s)\equiv {\bar h}^{a_1\ldots}_{b_1\ldots}(\{\lambda^{a_1\ldots}_{b_1\ldots}\};s) \, .
    \end{equation}
    Using these irreducible density matrices as generalized contractions, the basis of normal-ordered operators is defined recursively as
    \begin{subequations}
    \begin{align}
        :A^{a_1}_{b_1}: &= A^{a_1}_{b_1} - \lambda^{a_1}_{b_1}\,, \\
        :A^{a_1a_2}_{b_1b_2}: &= A^{a_1a_2}_{b_1b_2} - \mathcal{A}\left(\lambda^{a_1}_{b_1}\lambda^{a_2}_{b_2}\right) - \lambda^{a_1a_2}_{b_1b_2}\,, \\
        &\vdots\notag
    \end{align}
    \end{subequations}
    Details can be found in, e.g., Ref.~\cite{Hergert:2016etg}.

In actual calculations, the total energy is not strictly independent of $s$ due to the necessary approximations made on (i) the MR-IMSRG transformation that is truncated at the MR-IMSRG(2) level, i.e. mode-$2k$ tensors with $k\geq 3$ in Eq.~\eqref{IMsrgH} are consistently set to $0$ all throughout the transformation, and on (ii) the PGCM-PT expansion that is truncated at the PGCM-PT(2) level. Given the necessity to compute the PGCM-PT(2) correction anyway, one can avoid pushing the MR-IMSRG transformation to very large $s$ in order to limit the breaking of unitarity occurring at the MR-IMSRG(2) level.

\subsection{Numerical results}

The PGCM-PT(2) calculation of $^{20}$Ne is repeated for two MR-IMSRG(2) pre-processed Hamiltonians characterized by the flow parameters \(s = 10, 20\)\,MeV$^{-1}$ in addition to the unprocessed one ($s=0$) discussed earlier.  

\begin{figure}
    \centering
    \includegraphics[width=0.5\textwidth]{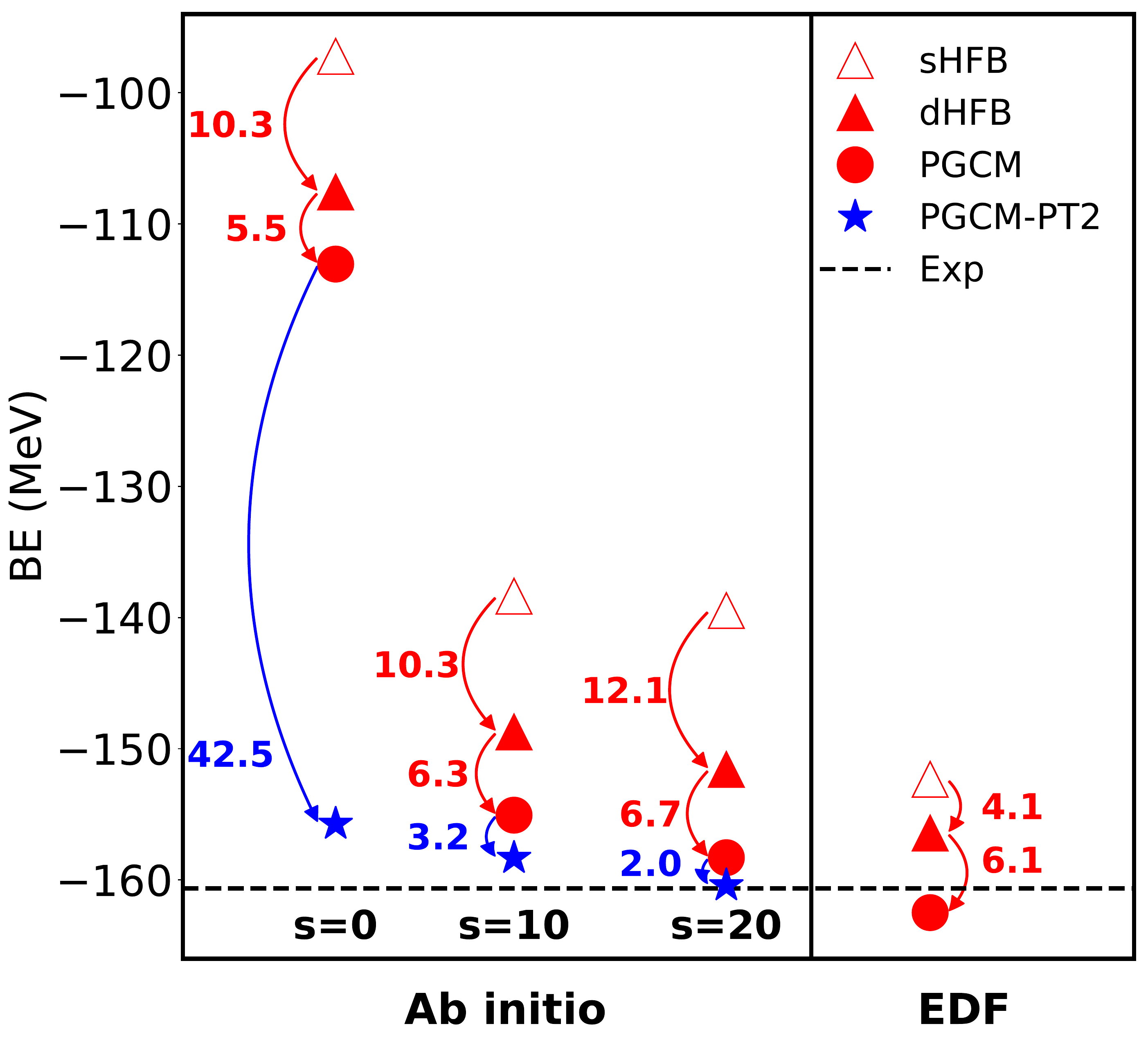}
    \caption{(color online) Same as Fig.~\ref{fig:ebran_plots} except that the ab initio calculation is performed for three values ($s=0,10,20$\,MeV$^{-1}$)  of the MR-IMSRG pre-processing parameter of the Hamiltonian.}
    \label{fig:ebran_plots_MRIMSRG}
\end{figure}

The systematic effect of the MR-IMSRG pre-processing of the Hamiltonian on the absolute binding energy of \nucl{Ne}{20} is shown in Fig.~\ref{fig:ebran_plots_MRIMSRG}. The MR-IMSRG evolution largely reshuffles the hierarchy of correlations at play. As $s$ grows, one observes that\footnote{The total energy is lowered by about 5\,MeV ($3\%$) from $s=0$\,MeV$^{-1}$ to $s=20$\,MeV$^{-1}$, i.e. it is not strictly invariant for reasons mentioned above. The problem becoming more perturbative with $s$ and the PGCM-PT(2) correction being reduced to $2$\,MeV at $s=20$\,MeV$^{-1}$, missing higher-order corrections are expected to be about few hundreds keV at that point. From the latter estimate and the variation of the total energy over the interval $s\in[0,20]$\,MeV$^{-1}$, the converged value can be estimated to lie within that 5\,MeV ($3\%$) band.}
\begin{enumerate}
\item the sHFB energy is drastically lowered due to the a prori resummation of the bulk of dynamical correlations into $\bar{H}(s)$,
\item static correlations captured at the PGCM level are stable throughout the process, only slightly increasing from about $16$\,MeV at $s=0$\,MeV$^{-1}$ to about $19$\,MeV at $s=20$\,MeV$^{-1}$, remaining overall consistent with the static correlations ($10$\,MeV) captured within the EDF calculation,
\item dynamical correlations explicitly grasped on top of PGCM through the PGCM-PT(2) correction are drastically reduced from being highly dominant at $s=0$\,MeV$^{-1}$ ($42$\,MeV) to being largely subleading at $s=20$\,MeV$^{-1}$ ($2$\,MeV).
\end{enumerate}
The picture that emerges is that, by resumming $95\%$ of the initial dynamical correlations into the pre-processed Hamiltonian $\bar{H}(s)$\footnote{The novel PGCM-PT formalism is instrumental to monitor the extent by which the MR-IMRG pre-processing can effectively reduce the need to resum dynamical correlations explicitly on top of the PGCM step, i.e., the extent by which the PGCM state is eventually decoupled from the rest of the Hilbert space as $s\rightarrow\infty$.}, the scales and hierarchy of correlations become highly consistent with those at play in the EDF calculation. Thus, the apparent inconsistency between the results obtained from both theoretical schemes in Sec.~\ref{sec:2} is lifted such that both a qualitative and quantitative link between them is now at reach.

\begin{figure}
    \centering
    \includegraphics[width=0.5\textwidth]{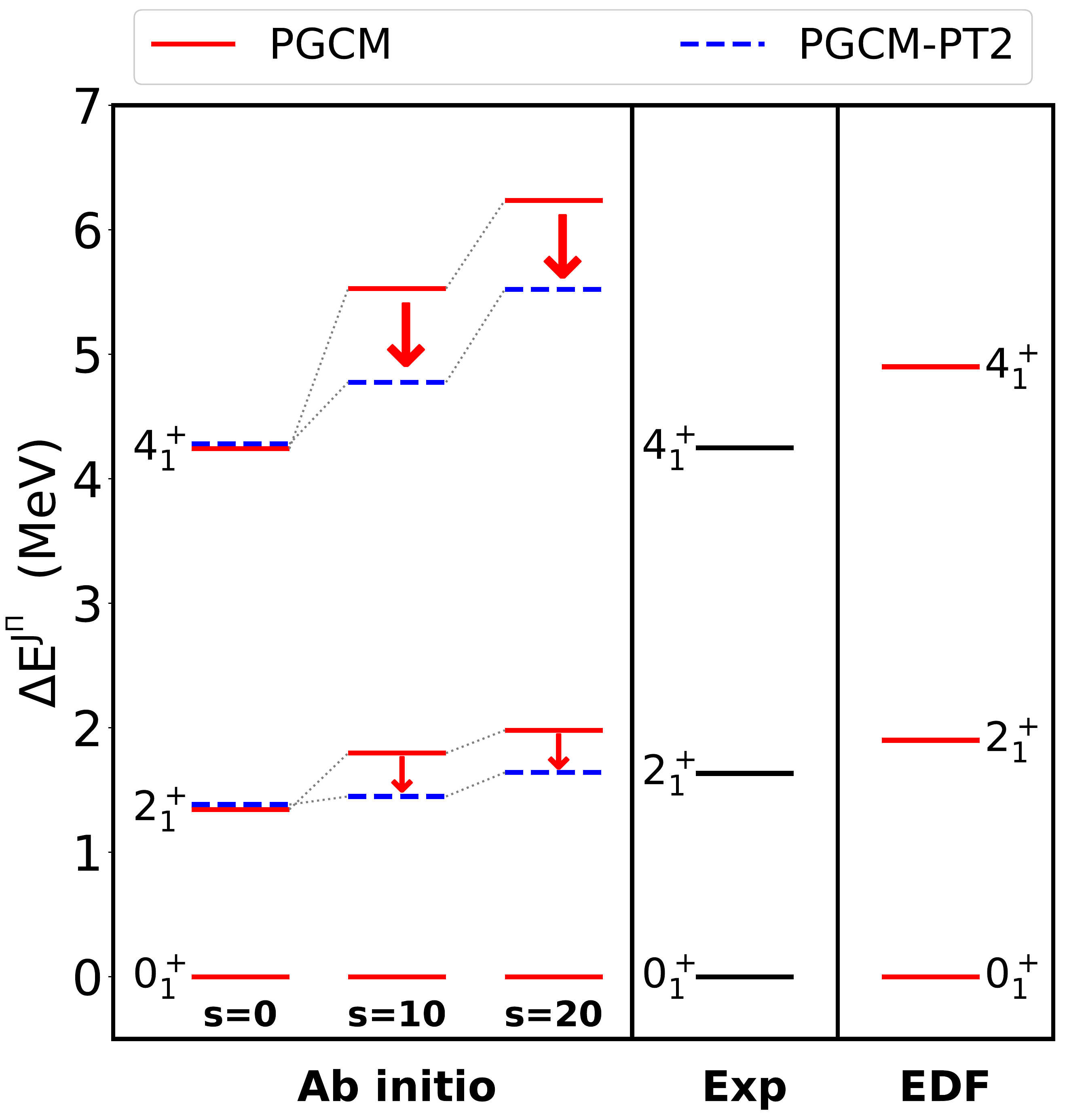}
    \caption{(color online) Same as Fig.~\ref{fig:ebran_plots_spectrum} except that the ab initio calculation is performed for three values ($s=0,10,20$\,MeV$^{-1}$) of the MR-IMSRG pre-processing parameter of the Hamiltonian.}
    \label{fig:ebran_plots_MRIMSRG_spectrum}
\end{figure}

The effect of the MR-IMSRG pre-processing on $^{20}$Ne ground-state rotational band is displayed in Fig.~\ref{fig:ebran_plots_MRIMSRG_spectrum}. As already seen in Fig.~\ref{fig:ebran_plots_spectrum}, the ab initio PGCM spectrum is essentially perfect at $s=0$\,MeV$^{-1}$ and the corresponding PGCM-PT(2) correction essentially zero. The picture changes significantly when pre-processing the Hamiltonian via MR-IMSRG. Indeed, the ab initio PGCM spectrum becomes more and more dilated as $s$ grows and resembles the EDF spectrum for $s \approx 10$\,MeV$^{-1}$. Consistently, the second-order PGCM-PT(2) correction increases with $s$ and systematically corrects for the dilatation, although not sufficiently for large $s$, in particular for the $4^{+}_1$ state.

\subsection{Discussion}

The above numerical results demonstrate that the empirical nuclear EDF method can be meaningfully rooted into the ab initio theoretical scheme for two key reasons.
\begin{enumerate}
    \item The family of MR-IMSRG pre-processed Hamiltonians $\bar{H}(s)$ is suited to play, for moderate-to-large $s$ values, the role of an effective and nucleus-dependent Hamiltonian acting within the restricted PGCM subspace ${\cal H}^{\text{PGCM}}_{\text{A}}$; i.e. the role of an ab initio version of $H_{\text{EDF}}$. Of course, using $\bar{H}(s)$ per se simply amounts to performing the ab initio calculation and to paying the computational cost of its construction in each nucleus, which does not fit the philosophy of an EDF-like scheme. As shown in Tab.~\ref{tab:complexity2}, building $\bar{H}(s)$ at the MR-IMSRG(2) level indeed scales as $O(n_{\text{dim}}^6)$ and thus largely dominates the cost of a PGCM calculation performed with it\footnote{When going to PGCM-PT(2), the construction of $\bar{H}(s)$ at the MR-IMSRG(2) level becomes subleading.}. Going to the MR-IMSRG(3) truncation scheme for greater accuracy further increases the scaling to $O(n_{\text{dim}}^9)$\footnote{The scaling of an MR-IMSRG(2) (MR-IMSRG(3)) calculation is presently given while excluding the use of the irreducible three-body density matrix (irreducible four- and five-body density matrices).}, which is already beyond current capabilities for state-of-the-art ab initio calculations~\cite{Heinz:2021xir,Hoppe:2021xqv}. In order to actually root the construction of $H_{\text{EDF}}$ into $\bar{H}(s)$, one must find ways to significantly decrease the cost to construct the latter. Possible lines of research in this direction are briefly alluded to in Sec.~\ref{sec:4} below.
    \item For moderate-to-large $s$ values, residual dynamical correlations captured via perturbation theory on top of PGCM are small. This is mandatory for a viable EDF-like scheme that wishes to effectively remain at the PGCM level and avoid altogether the need to perform, e.g., PGCM-PT(2) calculations whose cost scales as $O(n_{\text{dim}}^8)$. Strictly speaking though, the above results show that the residual coupling between the PGCM states and the complementary subspace is not zero for $s$ values of interest and even slightly magnified for the excitation spectrum compared to $s=0$\,MeV$^{-1}$. While an ab initio-rooted EDF scheme limited to a $O(n_{\text{dim}}^4)$ numerical scaling has to tolerate a degree of inaccuracy compared to a high-level ab initio scheme, it is of interest to further limit the coupling to the ${\cal Q}$ space for both ground and excited states at once. In this way, the systematic error made by omitting that coupling would be minimized. This idea will also be elaborated on Sec.~\ref{sec:4}.
\end{enumerate}

\begin{table}
    \centering
    \begin{tabular}{|l|c|c|}
    \hline
       Method  & MR-IMSRG(2) & MR-IMSRG(3) \\
           \hline
Runtime & \(O(n_{\text{dim}}^6)\) & \(O(n_{\text{dim}}^9)\)\\
Storage & \(O(n_{\text{dim}}^4)\) & \(O(n_{\text{dim}}^6)\) \\
           \hline
    \end{tabular}
    \caption{Runtime complexity and storage requirements of the MR-IMSRG(2) and MR-IMSRG(3) pre-processing methods.}
    \label{tab:complexity2} 
\end{table}

\section{Perspectives}
\label{sec:4}

As explained above, rooting an EDF-like scheme into the ab initio PGCM-PT formalism based on MR-IMSRG Hamiltonians requires (i) to make the PT corrections on top of the PGCM essentially negligible and (ii) to reduce significantly the cost to construct $\bar{H}(s)$. Indeed, this is mandatory to build an effective theory that operates with a mean-field-like scaling ($O(n_{\text{dim}}^4)$) independently of the nucleus under study and thus authorize a wide-spread or even universal applicability across the nuclear chart, ideally with a high level of physical interpretability. 

In the following, several potential lines of research to progess towards points (i) and (ii) are briefly elaborated on.

\subsection{Improving the ${\cal Q}$-space decoupling}
\label{sec:Qspace}

\subsubsection{Enriching the PGCM state}

The function of the MR-IMSRG pre-processing is to decouple the PGCM
state(s) from the rest of the Hilbert space and build the associated correlations into the evolved Hamiltonian. 

In practice, the span of ${\cal H}^{\text{PGCM}}_{\text{A}}$ can impact the quality of the decoupling. This is controlled by the choice of the generator coordinates actually defining the PGCM at play. The expectation is that progressively enriching the PGCM ansatz to include relevant relevant correlations can effectively reduce its coupling to the rest of the Hilbert space. 

The \nucl{Ne}{20} results displayed in Figs.~\ref{fig:ebran_plots} and \ref{fig:ebran_plots_MRIMSRG} were obtained with an ansatz based on axially and time-reversal symmetric fully paired HFB vacua. However, it was demonstrated earlier on~\cite{Frosini:2021sxj} that the inclusion of the octupole degree of freedom is necessary to describe the spectrum of this nucleus more quantitatively. In general, the inclusion of octupole, triaxial and/or pairing degrees of freedom is expected to be useful in a variety of doubly open-shell nuclei. Moreover, HFB states obtained via a variation after particle-number-projection calculation or via the use of a cranking constraint breaking time-reversal invariance can compress dilated PGCM spectra~\cite{PhysRevLett.113.162501,Borrajo:PLB2015,Yao:2019rck,Hergert20} and thus further suppress the residual dynamical correlations on top of the PGCM via, e.g. PGCM-PT. Last but not least, the inclusion of a few HFB states obtained via elementary excitations of fully paired HFB vacua is also expected to be useful in some specific situations~\cite{Frosini:2021sxj}, as a means to describe non-collective excitations. While the enrichment of the PGCM ansatz can significantly increase the CPU cost, it will remain highly subleading, especially as the nuclear mass grows, compared to the $n_{\text{dim}}^8$ cost\footnote{In the single-reference limit, second-order corrections are less costly and scale as $n_{\text{dim}}^5$. However, our objective is to design a the general framework that is multi-reference by nature and, as such, can address all nuclei.} associated with the explicit inclusion of dynamical correlations via PGCM-PT(2)\footnote{It is expected that this naive cost can however be significantly reduced in the future via optimization techniques~\cite{frosini2021}.}.

\subsubsection{Improving the MR-IMSRG transformation}

Apart from the characteristics of ${\cal H}^{\text{PGCM}}_{\text{A}}$, one may improve various characteristics of the MR-IMSRG transformation itself. While going to more advanced truncation schemes, e.g. MR-IMSRG(3), constitutes an obvious way to improve the $\mathcal{Q}$-space decoupling, it contradicts the goal to achieve the EDF-like $O(n_{\text{dim}}^{4})$ computational cost. It is thus not an option and other ideas must be envisioned. 

As already alluded to, the decoupling conditions are generally more complex for MR-IMSRG than for the SR-IMSRG: terms that vanish automatically for a Slater determinant reference state are no longer zero when normal-ordering the Hamiltonian with respect to PGCM states such that additional contributions originating from two-body and higher irreducible density matrices (cf. Eq. \eqref{eq:irred_density}) arise. Eventually, a state or a group of states cannot be \emph{strictly} decoupled from the rest of the Hilbert space. Still, approximate decoupling happens because the flow explores and eventually decouples the space of particle-hole like excitations that are built on the individual components of the correlated state \cite{Gebrerufael:2017fk,Yao:2018wq,Yao:2019rck}. Doing so, the evolution performed with respect to the PGCM ground state usually improves all states belonging to the associated rotational band although excited states tend to gain less energy because the transformation is (implicitly) tuned for the ground state reference. This feature likely explains the spreading of the excitation spectrum of \nucl{Ne}{20} in Sec. \ref{sec:3} and calls for an improvement.

\textit{Reference ensemble}

Starting with the latter point, the normal ordering and MR-IMSRG evolution can be performed with respect to a reference {\it ensemble} rather than with respect to a pure reference state. This procedure was first introduced in valence-space IMSRG (VS-IMSRG) calculations of open-shell nuclei \cite{Stroberg:2016ung,Stroberg:2019th} to account for partially filled orbitals in the calculation of ground-state energies. However, one can also view it as a mean to optimize the decoupling of the valence space as a whole. A similar observation was made in quantum chemistry \cite{Watson:2016eu}. Closer to the present problem of interest, ensembles of PGCM states were used to optimize the MR-IMSRG operator basis for the calculation of electromagnetic neutrinoless double beta decay transitions as well as the electromagnetic transitions in either nucleus \cite{Yao:2019rck,Yao:2021bs,Yao:2022nv}. Here, an ensemble built out of the PGCM states solution of the Hill-Wheeler equation could be employed to optimize the decoupling of  ${\cal H}^{\text{PGCM}}_{\text{A}}$ at once rather than of the ${\cal P}$ space defined by a single PGCM state. In this way, the entire PGCM spectrum could be improved in a single calculation\footnote{The costly alternative to improve the spectrum consists in performing one MR-IMSRG calculation per PGCM state. As far as the present discussion is concerned, this would correspond to generating one Hamiltonian $\bar{H}(s)$ per state, which is obviously not desirable.}, thus avoiding the spreading observed in Sec. \ref{sec:3}.

\textit{Improved operator basis}

It may be useful to see the MR-IMSRG through the lens of RG theory to tackle both accuracy and efficiency goals. The first step of any RG scheme is usually the selection of an appropriate operator basis to describe the RG flow. Importantly, the flow equations themselves can provide a diagnostic to identify relevant, marginal or irrelevant operators, and whether the initially chosen set is sufficiently complete whenever truncated or if relevant induced operators have been discarded. Furthermore, the identification and omission of irrelevant operators can immediately lead to a significant reduction of the computational effort whenever a high degree of redundancy exists in the operator basis. At the same time, the invariance of observables under MR-IMSRG evolution, or lack thereof, indicates whether the chosen operator basis is sufficiently complete and accurate. 

In the present context, the normal ordering procedure itself can be viewed as a mean to optimize
the basis of second-quantized operators $:A^{a_1\ldots}_{b_1\ldots}:$. As a matter of fact, going from simple reference states to suitably optimized correlated reference states, like PGCM states for open-shell nuclei, can turn some omitted operators from being relevant in SR-IMSRG(2) to being irrelevant in MR-IMSRG(2), thus reducing the truncation errors and flow-parameter dependence without altering the size of the basis. The optimal character of the operator basis can be further improved for a given reference state by exploiting the flexibility offered by the Mukherjee-Kutzelnigg scheme~\cite{Kong:2010kx}. First, it is possible to choose which of the irreducible density matrices are treated as contractions in Wick's theorem and which ones are included explicitly only when the many-body matrix element of an operator is evaluated. For example, the usual set of conditions
\begin{align}
        \bra{\Theta^{\sigma}_{\mu}}:A^{a_1\ldots}_{b_1\ldots}:\ket{\Theta^{\sigma}_{\mu}} = 0\,,
    \end{align}
    can be replaced by 
    \begin{subequations}
    \label{newcondition}
    \begin{align}
        \bra{\Theta^{\sigma}_{\mu}}:A^{a_1}_{b_1}:\ket{\Theta^{\sigma}_{\mu}} &= 0\,, \\
        \bra{\Theta^{\sigma}_{\mu}}:A^{a_1a_2}_{b_1b_2}:\ket{\Theta^{\sigma}_{\mu}} &= \lambda^{a_1a_2}_{b_1b_2} \,,\\
        \ldots{}\notag
    \end{align}
\end{subequations}
in order to adapt the definition of the normal-ordered operators at play. Indeed, there is evidence from applications in nuclear physics \cite{Hergert:2014iaa,Hergert:2016etg} and quantum chemistry~\cite{Datta:2011cr,Datta:2012zr} that it is sufficient to only retain the (correlated) one-body density matrix $\lambda^{a_1}_{b_1}$ of the reference state to construct normal-ordered operators according to Eq.~\eqref{newcondition}, thus making the tensors defining $\bar{H}(s)$ (Eq.~\eqref{IMsrgH}) functionals of the one-body density matrix alone. One must however be aware that the particular choice of generator and the ``extent'' of the transformation as measured by the flow parameter $s$ can affect whether the higher irreducible density matrices are indeed negligible. 

Second, the only property that the irreducible density matrices must actually fulfill to apply the normal ordering formalism is their antisymmetry under permutations of the upper and lower indices~\cite{Kong:2010kx}. Thus, it is in fact not necessary to require that the employed density matrices are associated with the reference, e.g. PGCM, state or with any many-body state for that matter. Relaxing this constraint and limiting Wick contractions to the one-body density matrix alone, the latter can be viewed as a tunable auxiliary quantity to be optimized for any given purpose\footnote{This idea was already employed in Ref.~\cite{Frosini:2021tuj} to optimize the rank-reduction of many-body operators.}, e.g. adapt the operator basis to improve the accuracy of the flow and to minimize the residual coupling to the rest of the Hilbert space. As a proof of principle, this idea was recently exploited to minimize the violation of unitarity of the MR-IMSRG flow in a pairing plus particle-hole interaction toy model \cite{Davison:2022a}. While this study relied on the ability to compute the exact eigenvalues of a complete block of $\bar{H}(s)$ in a full configuration approach, it can be adapted to realistic applications in the future. This will require the use of a cost function characterizing the unitarity of the evolution and/or the residual coupling while allowing a fast evaluation of the density matrix gradients, so that the desired $O(n_{\text{dim}}^4)$ is maintained. Eventually, the use of an auxiliary density matrix in the construction of the operator basis could also be useful to smooth out features of particular nuclei or states and help design a ``universal" EDF-like Hamiltonian $\bar{H}(s)$.

\subsection{Mitigating the construction cost of $\bar{H}(s)$}
\label{sec:h(s)}

One of the main reasons to pursue an approach with EDF-like $O(n_{\text{dim}}^4)$ scaling is that
will enable large-scale investigations of observables across the nuclear chart while also exploring the parameter spaces of the underlying interactions. This is key for the ongoing effort to refine our  understanding of these interactions and to systematically account for associated uncertainties.

While the previous section focused on ideas to avoid the $O(n_{\text{dim}}^8)$ cost of PGCM-PT(2) altogether, the present section looks into possibilities to reduce, or even bypass eventually, the MR-IMSRG cost associated with the construction of $\bar{H}(s)$, knowing that MR-IMSRG with explicit $k-$body operators, i.e., an MR-IMSRG(k) truncation, naively requires $O(n_{\text{dim}}^{3k})$ operations and $O(n_{\text{dim}}^{2k})$ memory.

Unsurprisingly, several efforts in the nuclear many-body community are underway to reduce these runtime and memory costs. At their heart lies the idea that these computational costs are in no small part caused by redundancies in the chosen representations of the interactions and many-body states, and that more efficient reduced-order models can be found (see Refs.~\cite{Melendez:2022kid,Bonilla:2022rph} for an introduction and overview).

\subsubsection{Reference ensemble}

One way to alleviate the construction cost of $\bar{H}(s)$ is to improve its universal character without yet reducing the cost of a single MR-IMSRG run. Naively, the MR-IMSRG evolution, and thus the construction of $\bar{H}(s)$, is to be performed for each quantum state of each nucleus. The idea discussed in the previous section to evolve the Hamiltonian with respect to the ensemble of PGCM states rather than with respect to each PGCM state separately is already a way to go beyond that naive approach. Furthermore, observables in the two nuclei involved in the neutrinoless double beta decay were shown in Ref.~\cite{Yao:2019rck,Yao:2021bs,Yao:2022nv} to be remarkably insensitive to sizable variations in the weights of the ensemble members. This observation is promising in view of yet enlarging the ensemble of PGCM states to generate MR-IMSRG evolved Hamiltonians $\bar{H}(s)$ that could not only be optimized for various quantum states of a given nucleus but also for several nuclei at once. It remains to be shown how far such an idea can pushed, i.e. what typical range of nuclei and quantum states can be successfully addressed from a single Hamiltonian $\bar{H}(s)$.

\subsubsection{Pre-processing techniques}

In recent years, different techniques aiming at reducing the computational cost of many-body methods have been put forward. The goal of such techniques is to lower the computation cost $O(n_{\text{dim}}^p)$ by (effectively) reducing either the exponent $p$ (e.g. via tensor factorization) or the dimension of the one-body Hilbert space $n_\text{dim}$ needed to yield converged results (e.g. via the use of importance truncation or natural orbitals).
While the former is expected to yield a more decisive advantage since it targets the exponent $p$ and thus directly meets the goal to go from, e.g. $O(n_{\text{dim}}^6)$ down to $O(n_{\text{dim}}^4)$, the combination of both types of pre-processing tools is likely to prove crucial in the long term in order to push the limits of applicability of many-body calculations.

\textit{Tensor factorization}

While (tensor) factorization techniques have been explored with great success in quantum chemistry \cite{Schutski:2017xr,Parrish:2019zg,Hohenstein:2022oj,Lesiuk:2020hq}, applications in nuclear physics are still in their infancy~\cite{Parrish:2013fe,Tichai:2019xz,Tichai:2022rq}.
The key challenge to achieve a scaling reduction is to reformulate many-body methods like the MR-IMSRG in such a way that they directly operate with the factors out of which the tensors defining the Hamiltonian and the unknown of the problem are built. This constitutes a promising but highly  nontrivial task~\cite{Tichai:2022rq,Schutski:2017xr,Parrish:2019zg}.

\textit{Importance truncation}

The goal of importance truncation (IT) techniques is to discard, prior to running the full costly calculation, some of the entries of the unknown many-body tensors at play in a given method.
This selection can be efficiently performed on the basis of a less expensive method, typically some variant of many-body perturbation theory. IT techniques have already been successfully applied to few many-body approaches~\cite{Roth:2009eu,TICHAI2019,Porro:2021rw} including the single-reference IMSRG~\cite{Hoppe:2021xqv}. 

\textit{Natural basis}

Among the possible strategies to optimise the single-particle basis, a promising procedure consists in the use of the so-called natural basis. The main idea is that natural orbitals, i.e. the eigenbasis of a correlated one-body density matrix, are already \textit{informed} from (dynamical) correlations at play, thus effectively reducing the size $n_\text{dim}$ of the one-body basis needed to explicitly incorporate those (dynamical) correlations via the high-level method of interest.
As for IT, the correlated one-body density matrix out of which natural orbitals are extracted is to be computed via some inexpensive approach like second-order perturbation theory. The use of natural orbitals has been recently explored in different methods~\cite{Tichai:2019to,Hoppe:2020elo,Fasano:2021ahd} such that the extension to MR-IMSRG can be envisaged in the future.

\subsubsection{Emulators}

In addition to accelerating individual MR-IMSRG runs needed to build $\bar{H}(s)$, emulators can be used to reduce the cost of large sets of calculations. They exploit smooth parameter dependences of the Hamiltonian, its eigenstates and associated observables. The full calculation is performed for a selection of parameter sets in order to generate training data for the emulator, which acts as a surrogate model that can predict the results for \emph{other} parameter values with high accuracy at a fraction of the computational cost.

The so-called eigenvector continuation (EC) approach constitutes a particularly successful 
basis for emulators of wave-function-based methods \cite{Frame:2018bv,Demol:2020nm,Konig:2020fn,Ekstrom:2019tw,Wesolowski:2021vr,Melendez:2021sy,Zhang:2022dk}. In the present context, a potentially impactful idea would be to treat the dependence of $\bar{H}(s)$ on the auxiliary density matrix via EC, so that MR-IMSRG evolutions would not have to be performed explicitly for each nucleus and parameter sets, but only for the training data. The main challenge is likely to be that the Hamiltonian depends on the density matrix and other parameters nonlinearly after the MR-IMSRG evolution whereas essentially all successful applications of EC so far have relied on linear parameter dependencies.

Naturally, it is also of great interest to attempt the emulation of the MR-IMSRG evolution itself. There, the goal is to emulate a \emph{unitary} transformation that acts directly on  the algebra of operators, without introducing the A-body Hilbert space representation that leads to dealing with large matrix representations of these evolving operators. This rules out the use of the wave-function-based EC such that other avenues for model-order reduction must be explored. Research in this direction is in progress \cite{Davison:2022b}.

It must be remarked that the quality of any emulator necessarily relies on the availability of sufficient training data, especially if the parameter dependence is high dimensional. Thus, new developments for emulators must go hand in hand with the efficiency improvements of individual MR-IMSRG runs, so that the training data needs can be readily met.

\subsection{Towards parameterizations of $\bar{H}(s)$}
\label{sec:H_param}

An important research track that complements the work described in the previous sections
is the construction of explicit parameterizations of $\bar{H}(s)$ that would eventually bypass the need to actually run any MR-IMSRG calculation. 

In the near term, the empirical dependence of its zero-, one- and two-body components on proton and neutron numbers, or even on the one-body density matrix can be envisioned. As discussed earlier on, an auxiliary density matrix can be used in order to generate smooth trends, even if this comes at the cost of a slightly reduced accuracy for certain
observables. 

In the longer term, the empirical information accumulated through the aforementioned investigations can help developing an EFT formulated with, e.g., the one-body density matrix as a degree of freedom to directly construct $\bar{H}(s)$ in a systematic fashion without ever resorting to an explicit MR-IMSRG transformation of the Hamiltonian at play in the underlying chiral effective field theory. 

\section{Conclusions}
\label{conclusions}

The present paper demonstrates, based on both formal arguments and numerical results, that the empirical nuclear energy density functional (EDF) method can be convincingly rooted into the recently formulated ab initio many-body perturbation theory built on top of the projected generator coordinate method (PGCM-PT), whenever the latter employs an effective Hamiltonian resulting from a multi-reference in-medium similarity renormalization group (MR-IMSRG) transformation of the nuclear Hamiltonian at play in chiral effective field theory. Indeed, such a pre-processed Hamiltonian effectively resums dynamical correlations, making perturbative corrections on top of the PGCM eventually negligible. Consequently, the PGCM states at play in the EDF method end up being good approximations to low-lying eigenstates of the pre-processed Hamiltonian that, in turn, can be seen as the ab initio counterpart of the effective nucleus-dependent Hamiltonian at play in the empirical EDF method.

Based on the above analysis, the second objective of the present work is to discuss possible lines of research leading to the design of an ab initio-rooted nuclear EDF method. Given that the fundamental benefit of the (empirical) EDF method lies in its mean-field-like $O(n_{\text{dim}}^4)$ scaling authorizing large-scale studies across the nuclear chart, the main challenges to achieve this goal are chiefly computational. Thus, ideas have been put forward to
\begin{itemize}
\item further reduce the size of dynamical corrections on top of the PGCM such that one can effectively remain at the latter level at the price of an acceptable error, in order to avoid altogether the expensive, e.g. $O(n_{\text{dim}}^8)$ at the PGCM-PT(2) level, calculation needed to grasp them on top of the PGCM,
\item alleviate the $O(n_{\text{dim}}^6)$ ($O(n_{\text{dim}}^9)$) cost of the effective Hamiltonian construction at the MR-IMSRG(2) (MR-IMSRG(3)) level.
\end{itemize}
Based on the know-how collected via the realization of the above objectives, one could contemplate the idea to eventually develop an EFT formulated with, e.g., the one-body density matrix as a degree of freedom to directly construct the effective EDF Hamiltonian in a systematic fashion without ever resorting to an explicit MR-IMSRG transformation of the Hamiltonian at play in the underlying chiral effective field theory. 

\section*{Acknowledgements}

Some of the calculations presented here were performed by using HPC resources from GENCI-TGCC, France (Contract No. A0110513012).

\bibliography{bibliography.bib}

\end{document}